\documentclass[useAMS,usenatbib]{mn2e}
\usepackage{graphicx,natbib,times}

\title[`Hanny's Voorwerp', a quasar light echo?]{Galaxy Zoo: `Hanny's Voorwerp', a quasar light echo?\thanks{This publication has been made possible by the participation of more than 100,000 volunteers in the Galaxy Zoo project. Their contributions are individually acknowledged at http://www.galaxyzoo.org/Volunteers.aspx}}
\author[Lintott et al.]{
  \parbox[t]{16cm}{Chris J. Lintott$^{1}$\thanks{E-mail: cjl@astro.ox.ac.uk (CJL)},
    Kevin Schawinski$^{1,2,3}$, William Keel$^{4,5}$\thanks{Visiting Astronomer, Kitt Peak National Observatory, National Optical Astronomy Observatory, which is operated by the Association of Universities for Research in Astronomy (AURA) under cooperative agreement with the National Science Foundation.}, Hanny van Arkel$^{6}$,  Nicola Bennert$^{7,8}$, Edward Edmondson$^{9}$, Daniel Thomas$^{9}$,  Daniel J.B. Smith$^{10}$, Peter D. Herbert$^{11}$, Matt J. Jarvis$^{11}$, Shanil Virani$^{3}$,Dan Andreescu$^{12}$, Steven P. Bamford$^{8}$, Kate Land$^{1}$, Phil Murray$^{13}$, Robert C. Nichol$^{8}$, M. Jordan Raddick$^{14}$, An\v{z}e Slosar$^{15}$, Alex Szalay$^{14}$, Jan Vandenberg$^{14}$\\ 
  }\\
$^{1}$Department of Physics, University of Oxford,
Oxford OX1 3RH, UK.\\
$^{2}$ Department of Physics, Yale University, New Haven, CT 06511, USA\\
$^{3}$ Yale Center for Astronomy and Astrophysics, Yale University, P.O. Box 208121, New Haven, CT 06520, USA\\
$^{4}$ Dept. of Physics and Astronomy, University of Alabama,
Box 870324, Tuscaloosa, AL 35487, USA.\\
$^{5}$ SARA Observatory\\
$^{6}$ Netherlands School System\\
$^{7}$ Institute of Geophysics and Planetary Physics, University of California, Riverside, CA 92521\\
$^{8}$ Physics Department, University of California, Santa Barbara, CA 93106, USA.\\
$^{9}$ Institute of Cosmology \& Gravitation, University of Portsmouth, UK.\\
$^{10}$ Astrophysics Research Institute, Liverpool John Moores University, Twelve Quays House Egerton Wharf, Birkenhead, CH41 1LD, UK.\\
$^{11}$ Centre for Astrophysics, Science \& Technology Research Institute, University of Hertfordshire, Hatfield, UK.\\
$^{12}$ LinkLab, 4506 Graystone Ave., Bronx, NY 10471, USA.\\
$^{13}$Fingerprint Digital Media, 9 Victoria Close, Newtownards, Co. Down, Northern Ireland, BT23 7GY, UK.\\
$^{14}$Department of Physics and Astronomy, Johns Hopkins University, 3400 N. Charles St., Baltimore, MD 21218, USA.\\
$^{15}$Berkeley Centre for Cosmological Physics, Lawrence Berkeley National Laboratory and Physics Department, Berkeley, CA 94720, USA.\\
}

\begin{document}

\newcommand\aj{{AJ}}%
\newcommand\actaa{{Acta Astron.}}%
\newcommand\araa{{ARA\&A}}%
\newcommand\apj{{ApJ}}%
\newcommand\apjl{{ApJ}}%
\newcommand\apjs{{ApJS}}%
\newcommand\ao{{Appl.~Opt.}}%
\newcommand\apss{{Ap\&SS}}%
\newcommand\aap{{A\&A}}%
\newcommand\aapr{{A\&A~Rev.}}%
\newcommand\aaps{{A\&AS}}%
\newcommand\azh{{AZh}}%
\newcommand\baas{{BAAS}}%
\newcommand\caa{{Chinese Astron. Astrophys.}}%
\newcommand\cjaa{{Chinese J. Astron. Astrophys.}}%
\newcommand\icarus{{Icarus}}%
\newcommand\jcap{{J. Cosmology Astropart. Phys.}}%
\newcommand\jrasc{{JRASC}}%
\newcommand\memras{{MmRAS}}%
\newcommand\mnras{{MNRAS}}%
\newcommand\na{{New A}}%
\newcommand\nar{{New A Rev.}}%
\newcommand\pra{{Phys.~Rev.~A}}%
\newcommand\prb{{Phys.~Rev.~B}}%
\newcommand\prc{{Phys.~Rev.~C}}%
\newcommand\prd{{Phys.~Rev.~D}}%
\newcommand\pre{{Phys.~Rev.~E}}%
\newcommand\prl{{Phys.~Rev.~Lett.}}%
\newcommand\pasa{{PASA}}%
\newcommand\pasp{{PASP}}%
\newcommand\pasj{{PASJ}}%
\newcommand\qjras{{QJRAS}}%
\newcommand\rmxaa{{Rev. Mexicana Astron. Astrofis.}}%
\newcommand\skytel{{S\&T}}%
\newcommand\solphys{{Sol.~Phys.}}%
\newcommand\sovast{{Soviet~Ast.}}%
\newcommand\ssr{{Space~Sci.~Rev.}}%
\newcommand\zap{{ZAp}}%
\newcommand\nat{{Nature}}%
\newcommand\iaucirc{{IAU~Circ.}}%
\newcommand\aplett{{Astrophys.~Lett.}}%
\newcommand\apspr{{Astrophys.~Space~Phys.~Res.}}%
\newcommand\bain{{Bull.~Astron.~Inst.~Netherlands}}%
\newcommand\fcp{{Fund.~Cosmic~Phys.}}%
\newcommand\gca{{Geochim.~Cosmochim.~Acta}}%
\newcommand\grl{{Geophys.~Res.~Lett.}}%
\newcommand\jcp{{J.~Chem.~Phys.}}%
\newcommand\jgr{{J.~Geophys.~Res.}}%
\newcommand\jqsrt{{J.~Quant.~Spec.~Radiat.~Transf.}}%
\newcommand\memsai{{Mem.~Soc.~Astron.~Italiana}}%
\newcommand\nphysa{{Nucl.~Phys.~A}}%
\newcommand\physrep{{Phys.~Rep.}}%
\newcommand\physscr{{Phys.~Scr}}%
\newcommand\planss{{Planet.~Space~Sci.}}%
\newcommand\procspie{{Proc.~SPIE}}%
\newcommand\helvet{{Helvetica~Phys.~Acta}}%

\date{October 2008}

\pagerange{\pageref{firstpage}--\pageref{lastpage}} \pubyear{2008}

\maketitle

\label{firstpage}

\begin{abstract}
We report the discovery of an unusual object near the spiral galaxy IC 2497, discovered by visual inspection of the Sloan Digital Sky Survey (SDSS) as part of the Galaxy Zoo project. The object, known as Hanny's Voorwerp, is bright in the SDSS $g$ band due to unusually strong [O\textsc{iii}] 4959,~5007 emission lines. We present the results of the first targeted observations of the object in the optical, UV and X-ray, which show that the object contains highly ionized gas. Although the line ratios are similar to extended emission-line regions near luminous AGN, the source of this ionization is not apparent. The emission-line properties, and  lack of x-ray emission from IC 2497, suggest either a highly obscured AGN with a novel geometry arranged to allow photoionization of the object but not the galaxy's own circumnuclear gas, or, as we argue, the first detection of a quasar light echo. In this case, either the luminosity of the central source has decreased dramatically or else the obscuration in the system has increased within $10^5$ years. This object may thus represent the first direct probe of quasar history on these timescales.
\end{abstract}


\section{Introduction}
The Galaxy Zoo project\footnote{www.galaxyzoo.org} (Lintott et al. 2008) has completed a morphological classification of almost 900,000 objects drawn from the Sloan Digital Sky
Survey (SDSS; \citealt{York, DR6}). By combining classifications made by more than 100,000 participants, it proved possible to compile catalogues of morphology which are of comparable accuracy to those produced by professional astronomers, despite being an order of magnitude larger. The data produced were primarily intended for use in the study of the properties of the population of galaxies (e.g. Bamford et al. 2008), but visual inspection of images from surveys such as the SDSS provides an excellent way of identifying unusual objects within the data set. 

In this paper, we discuss an unusual structure, colloquially known as `Hanny's Voorwerp'\footnote{`Voorwerp' is Dutch for object.} discovered by Hanny van Arkel in the vicinity of the spiral galaxy IC 2497. We report this discovery and present the results of initial follow-up observations in the visible, ultraviolet and X-ray regions of the spectrum. We consider the emission-line spectrum in
detail, and consider possible sources for the observed degree of ionization.

\section{Pre-existing observations of IC 2497}
\label{sec:IC2497}

While there are no pre-existing observations of our target, the neighbouring galaxy IC 2497, is included in several surveys. It has a measured redshift of $\rm z
= 0.050221$\footnote{NASA/IPAC Extragalactic Database,
\texttt{http://nedwww.ipac.caltech.edu/}} \citep{Fisher}. Assuming, as we will throughout this paper, $\mathrm{H_0}=71$, $\Omega_m=0.27$ and $\Omega_\lambda=0.73$ \citep{Dunkley} this redshift corresponds to a luminosity distance of 220.4 Mpc and a scale of 969 $\mathrm{pc\,arcsec^{-1}}$. With an absolute magnitude
of $M_{r} = -22.1~\rm mag$ it is a luminous system around 1.7 mags brighter than  $M_{*,r}$
\citep{Blanton}. The SDSS imaging shows it as a disk galaxy with a large bulge and two fainter spiral arms, as shown in Figure \ref{fig:sdss_voorwerp}. 
IC 2497 is also detected at radio
wavelengths in the Very Large Array (VLA) FIRST survey
\citep{Becker}, with a flux 16.1$\pm0.8$ mJy at 1.4 GHz, and hence a radio luminosity of $L_{1.4~\rm GHz}
= 1.00\pm 0.05 \times 10^{23}~ \rm WHz^{-1} $. 

IC 2497 was also detected by
\textit{IRAS} (the Infrared Astronomical Satellite) at 25, 60, and 100$\mu
m$, with values from the point-source catalogue giving it an infrared luminosity 
of $L_{\rm IR} = 3.9 \times
10^{11} L_{\odot}$ \citep{SandersMirabel}, and is thus a Luminous InfraRed Galaxy (LIRG). However, inspection of the IRAS data using the Infrared Sky Atlas (IRAS) tool at the Infrared Processing and Analysis Centre (IPAC\footnote{http://www.ipac.caltech.edu}) web archive shows that the 60$\mu$m measurement (and possibly the others) may be confused with a stronger source about 2 arcminutes to its south.

To verify the IRAS fluxes, we used the SCANPI web tool from IPAC to retrieve fluxes for each
detector crossing of IC 2497, establishing the absence of confusing
sources and averaging the scans for measurement. The resulting flux
densities were 0.14, 0.22, 2.04, and 3.71 Jy in the 12, 25, 60, and
100$\mu$m bands respectively, with errors of 0.02, 0.02, 0.02, and 0.06 Jy.
Using the FIR parameter from Lonsdale et al. (1985)
 the luminosity from 42-122$\mu$m is $6 \times 10^{44}$ erg s
$^{-1}$. Despite the
high luminosity for such an ordinary-looking galaxy, we note that the
FIR energy distribution suggests emission from a source which is colder than most AGN-dominated sources.

\section{Imaging Data}

\subsection{SDSS Imaging Data \& Photometric Properties}
Hanny's Voorwerp was initially identified in visual inspection of SDSS imaging. In Figure
\ref{fig:sdss_voorwerp}, we present the full SDSS $ugriz$ imaging including
a $gri$ colour composite \citep{Lupton} similar to that displayed by the Galaxy Zoo website. 
 
The SDSS photometric data clearly flag the Voorwerp as an unusual
object. It is a significant detection only in the $g$-band, where it reaches
an apparent magnitude of $g=18.84$.  Integrated magnitudes within an aperture of 10 arcsec in the SDSS bands (but in the Pogson logarithmic convention rather than the SDSS sinh style) are $u=20.5\pm0.15$, $g=18.12\pm0.08$, $r=21.3\pm0.1$ and $i=19.8\pm0.1$. The object was not detected in the $z$-band.  Similar distributions between bands are seen at each of the six SDSS photometric objects associated with the Voorwerp, justifying the use of integrated magnitudes. The most unusual is the `knot' to the NW, which has a different spectral energy distribution from the bulk of the object, being particularly bright in the $r$ and $i$ bands. This may suggest contamination by a background source, but a spectrum of the knot itself would be required to test this hypothesis. It is relatively unimportant in the $g$ band, contributing less than 15\% of the integrated flux.

The unusual colours of the Voorwerp itself result both from the strength of [O III]$\lambda\lambda$4959,5007\AA~ and the fact that this redshift places H$\alpha$ on the red wing of the $r$ passband response. This is clearly an unusual object, and worthy of further study.

\subsection{INT Imaging Data}
\begin{figure*}
\begin{center}

\includegraphics[angle=90, width=\textwidth]{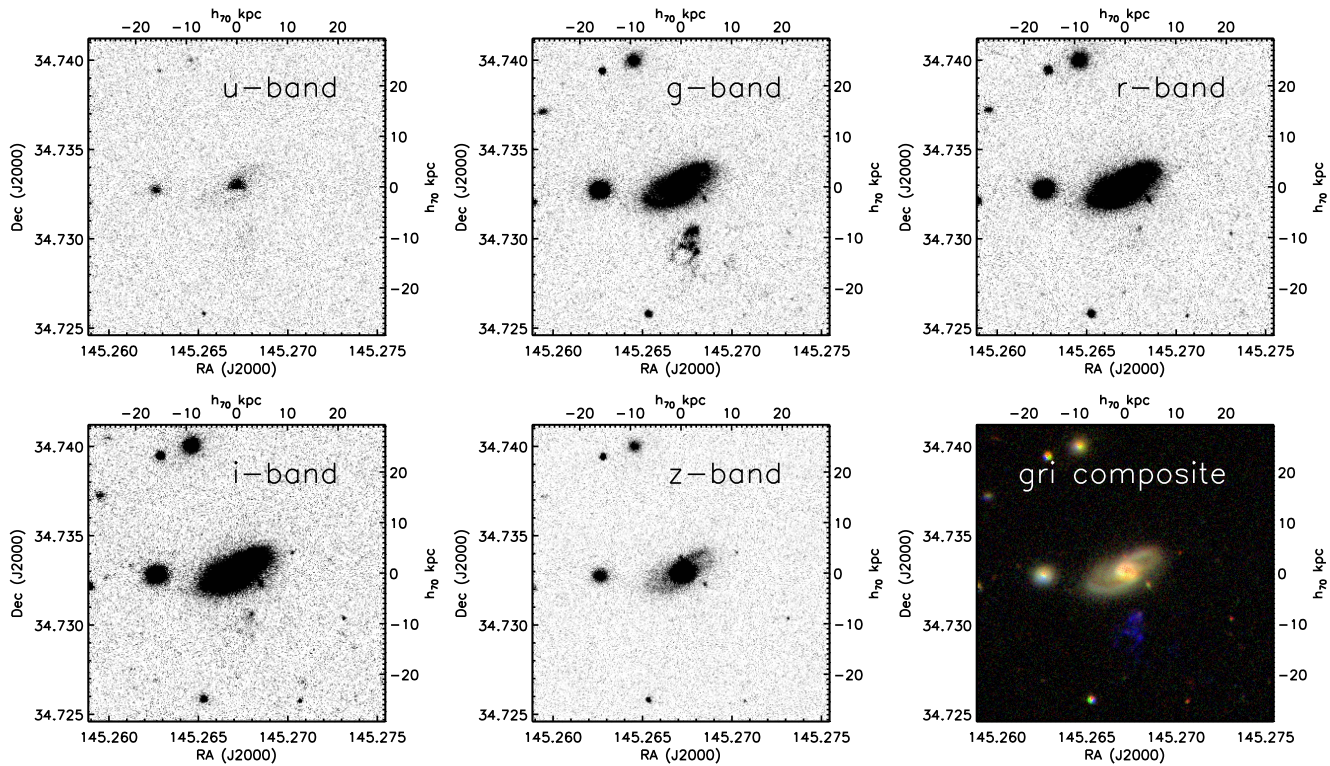}
\caption{The SDSS images of the main galaxy IC 2497 and the
Voorwerp. We show each of the five SDSS images ($u$, $g$, $r$, $i$ and
$z$ separately as well as a three-colour $gri$ composite
\citep{Lupton}. The latter is similar to the Galaxy Zoo
image in which the Voorwerp was discovered. IC 2497 is clearly visible
in all five bands, while the Voorwerp is strikingly prominent only in
the $g$-band. It is marginally detected in the $u$, $r$, and $i$-bands, and
undetected in $z$. We also indicate the physical scale in
$\rm h_{70} kpc$ at the redshift of the system. \label{fig:sdss_voorwerp}}
\end{center}
\end{figure*}

\begin{figure*}
\begin{center}

\includegraphics[angle=90, width=0.48\textwidth]{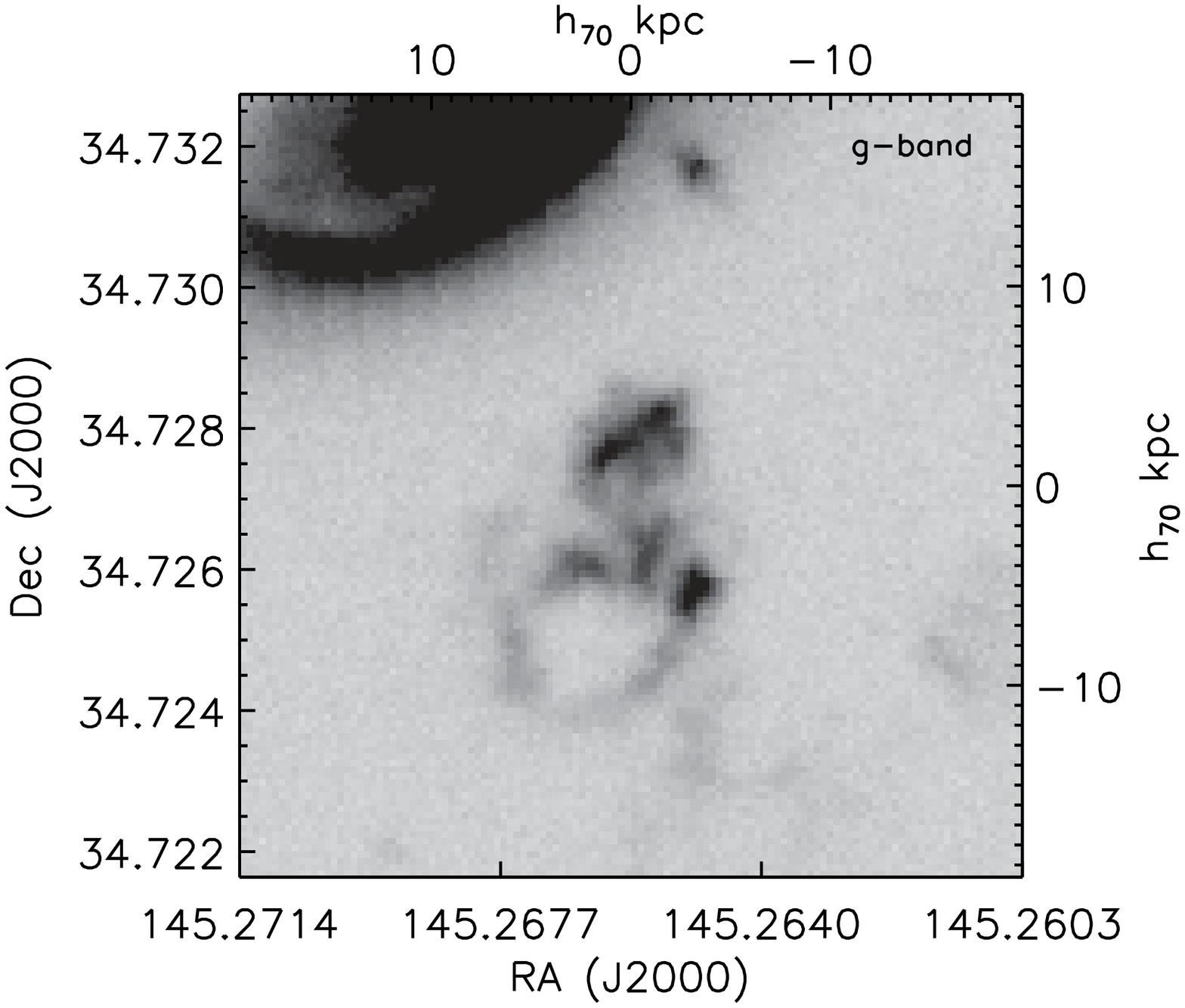}
\includegraphics[angle=90, width=0.48\textwidth]{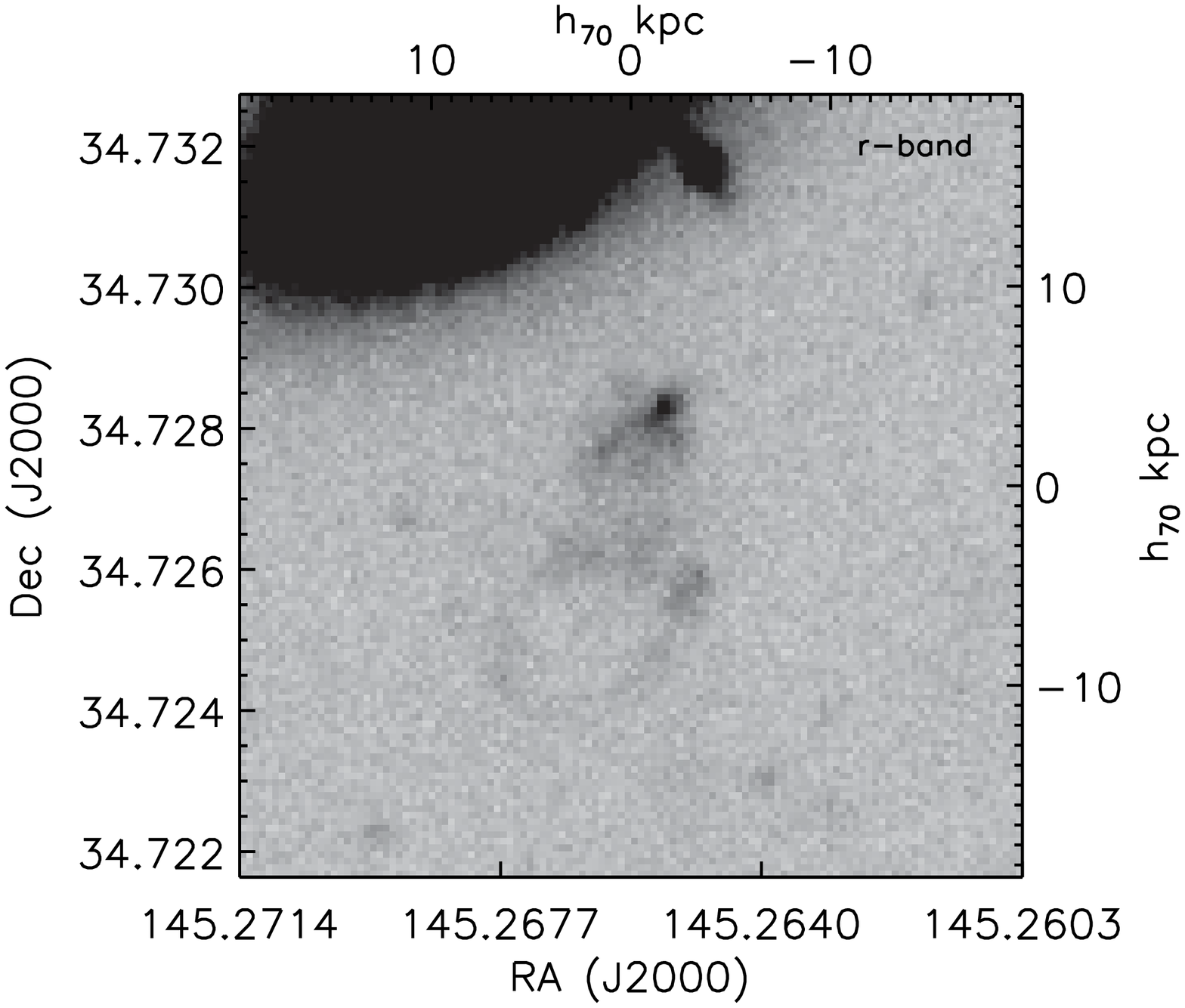}
\includegraphics[angle=90, width=0.48\textwidth]{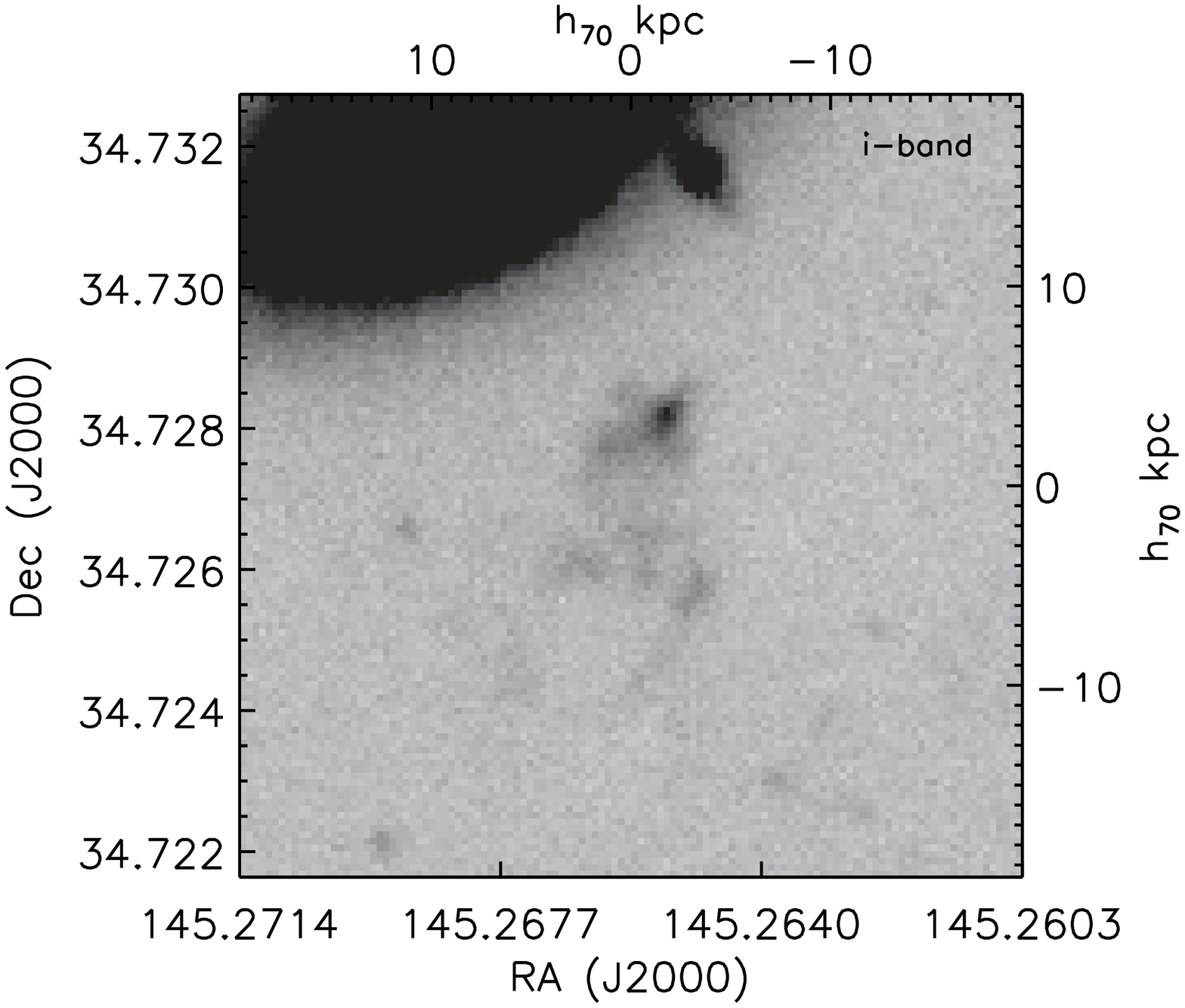}
\includegraphics[angle=90, width=0.48\textwidth]{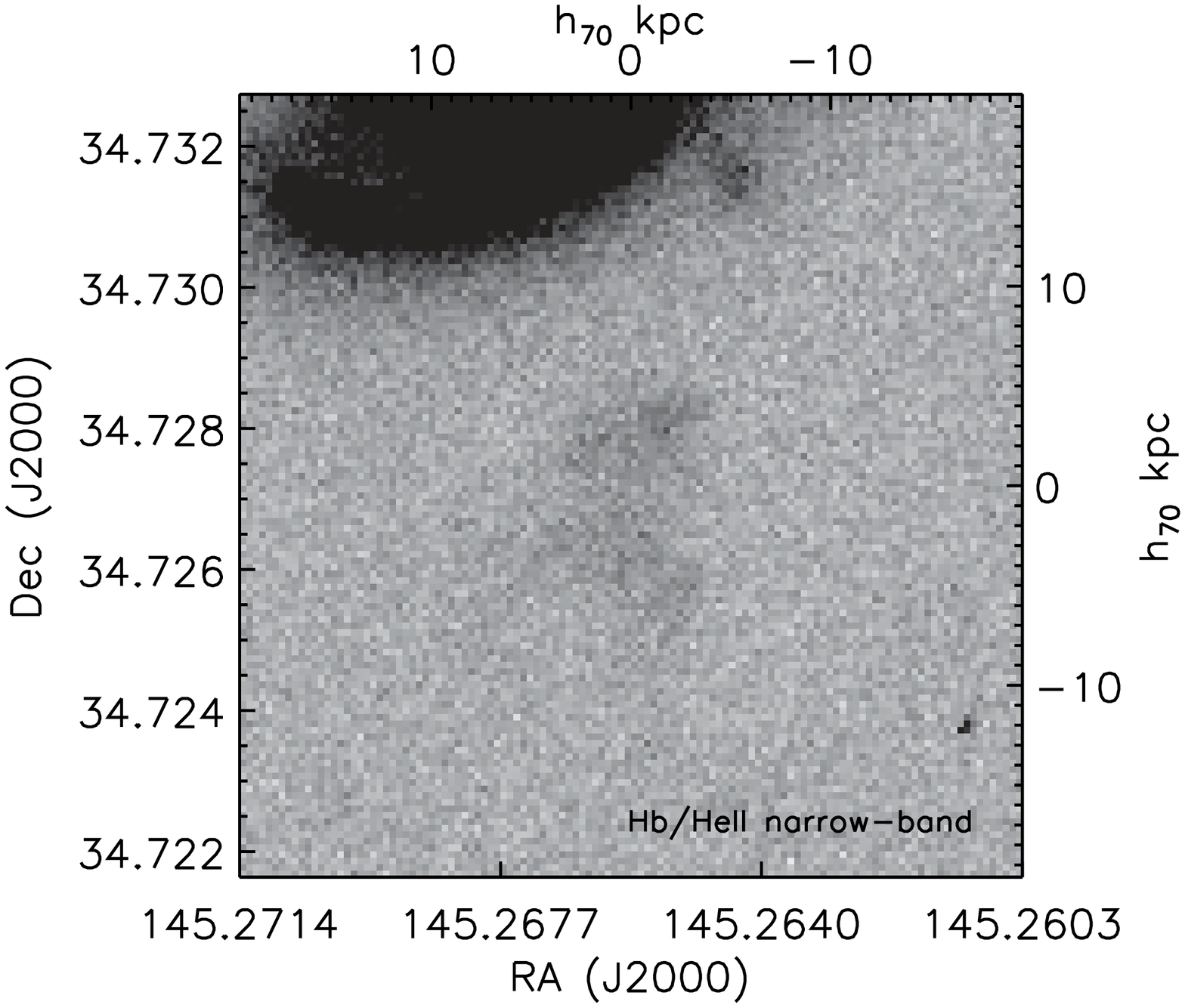}
\caption{The INT images of the Voorwerp. These $g$, $r$ and $i$-band images are significantly deeper
than the SDSS images in Figure \ref{fig:sdss_voorwerp} and were
obtained in better seeing conditions. We also present a narrow-band
image in the broad H$\beta$ filter (centred on $\lambda = 4861\rm\AA$;
FWHM = 170$\rm\AA$), which at the redshift of IC 2497 and the
Voorwerp traces HeII $\lambda 4686$. We also indicate the physical scale in
$\rm h{70} kpc$ at the redshift of the system.\label{fig:int_voorwerp}}

\end{center}
\end{figure*}

We have obtained a series of deeper imaging data from the Wide-Field
Imager (WFI) at the Isaac Newton Telescope (INT). The data consist of
three 400s images in each of the $g$, $r$ and $i$ bands (on 2008 January 11) and, on the 2008 January 9, a 600s
image using the wide H$\beta$ narrow-band filter (centred on $\lambda
= 4861\rm\AA$; FWHM = 170$\rm\AA$), which at the redshift of IC 2497
traces HeII $\lambda 4686$. All four images are shown in Figure
\ref{fig:int_voorwerp}. The $g$-band image, deeper than that in SDSS, reveals that the
Voorwerp is a significantly larger system than was previously apparent in the data shown in Figure \ref{fig:sdss_voorwerp}. The detected emission extends over 18$\times$40'' (EW vs NS) with additional outlying emission visible to the west. 

The morphology of the object is complex, and includes several prominent features. The $g$-band images, which are dominated by [O\textsc{iii}] 4959, 5007 emission, reveal a lumpy structure, particularly in the part of the object closest to IC 2497. Moving further away, several smaller discrete structures appear that form a nearly round `bubble'. This hole is 5.4 arcsec in diameter (corresponding to 4.9~kpc at the distance of IC 2497). The high degree
of symmetry seen in this structure poses puzzling questions about its
origin.

Remarkably, the Voorwerp is detected in the He\textsc{ii} narrow-band image with the He\textsc{ii} emission in this band coinciding with the brightest features seen in the $g$-band.


\subsection{H$\alpha$ Imaging Data}

\begin{figure}
\includegraphics[angle=90, width=0.49\textwidth]{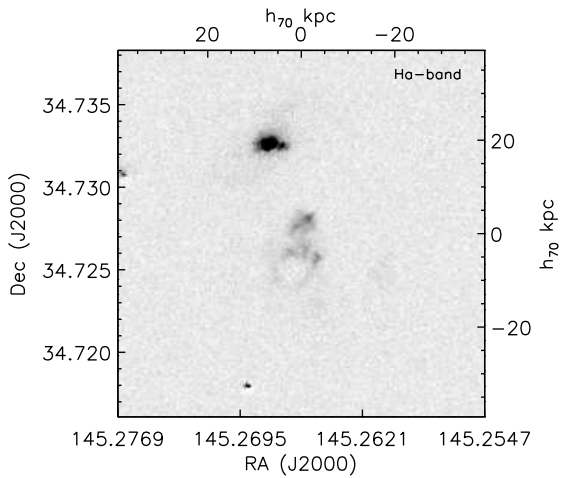}
\caption{H$\alpha$ emission from the Voorwerp field observed with the KPNO 2.1m. The Voorwerp itself is prominent. We note also the emission source to the SW of the IC 2497 nucleus, which is not seen so clearly in any other band.}\label{fig:halpha}
\end{figure}

We also obtained an image of the field through a filter centred on the H$\alpha$ line, as well as an off-line exposure on the adjacent continuum. These images were taken with the Kitt Peak 2.1m telescope on 2008 March 27, using a 2k x 2k TI CCD which sampled the image with 0.305~arcsec $\mathrm{pix}^{-1}$. Exposures were 30 minutes each in filters centred at observed wavelengths of 6877 and 6573 \AA, with spectral FWHM 76 and 69 \AA~ respectively. Star images showed FWHM of 0.85 arcsec. Even with smoothing and integration across the extent of the object, the continuum (off-line) image shows little flux at the Voorwerp's position; at these wavelengths emission from the object is thus strongly dominated by the emission lines. Flux calibration of these images was accomplished through observations of the
standard star Feige 15 and the planetary nebula NGC 2392 (with integrated fluxes from \citet{Pottasch}) in three matched filters with neighbouring
passbands. These two standards are complementary in that the star has strong signal in all filters, while the nebula measures do not depend on accurate knowledge of the filter width. Results from these two standards agree within 3\% in intensity scale. A net emission-line image was produced and is shown in Figure \ref{fig:halpha}.

Both IC 2497 and the Voorwerp are clearly seen in the difference image.  The characteristic shape of the Voorwerp, including the `bubble' discussed above, is clearly seen. We also note the appearance of a second, resolved source 2.3" to the WSW of the IC 2497 nucleus, suggestive of a double nucleus or a minor merging event. While this is most clearly seen in the H$\alpha$ image, this source is present in our optical continuum images, showing that it is substantially a continuum object.

\subsection{Deep continuum imaging in $R$}

To provide a better measurement of the red continuum, a total exposure of 110 minutes in the Bessel $R$ band was obtained on 2008 April 27/28, using the remotely-operated 0.9m telescope of the Southeastern Association for Research in Astronomy (SARA) sited on Kitt Peak. The detector was a $2048 \times 2048$ pixel E2V chip in an Apogee U42 camera, giving pixel sampling of 0.38 arcsec $\mathrm{pix^{-1}}$. The passband used has H$\alpha$ and [N II]$\lambda$6583\AA~ in the red wings of its transmission, so correction for their effects introduces only a small uncertainty. Using the energy zero points from \citet{Fukugita} and the same integration region used for total flux from the INT $g$ image, we derive an averaged flux in $R$ across the emitted-wavelength range 5900-6500 \AA~ of $8.8 \pm 1.0 \times 10^{-18}$ erg cm$^{-2}$ 
s$^{-1}$ \AA$^{-1}$.

\section{Spectral data}
\label{sec:spec}
Spectra covering most of the optical band were obtained with double-spectrograph systems at the
4.2m William Herschel Telescope (WHT) on La Palma and the 3m Shane telescope of Lick Observatory.
Details of the observations are given in Table \ref{tab:spectra}. The slit width was 2.0 arcsec in both cases, and placement on the sky was nearly identical, passing in both cases through the nucleus of IC 2497. 

We applied the same reduction procedure to each data set. To eliminate the ripples in sensitivity due to the dichroic beamsplitters in each double spectrograph, which are especially troublesome near [O III]$\lambda$ 5007 at this redshift, we used the flat-field exposures as obtained, omitting the common step of removing large-scale spectral gradients. After flat-fielding, the spectra thus appeared very blue, but the response curves generated from standard stars were monotonic across almost the entire spectral range and were well fitted by  smooth functions. The region containing  H$\beta$ and [O III]$\lambda\lambda$ 4959, 5007 falls very close to the rollover wavelength for each dichroic at this redshift, and the derived line ratio is thus very sensitive to how well these transmission ripples can be corrected.

\begin{table*}
\begin{tabular}{|l|c|c|c|c|}
\hline
 Telescope                    &        WHT 4.2m	& Lick 3m		& Lick 3m \\ 
Spectrograph                  &   ISIS		& Kast			&Kast \\
Exposure (minutes)         &  30		& 30			&2$\times$x30 \\
PA (degrees)                      & 9.5		& 8.5			& 8.5 \\
Slit width (arcsec)              & 1.97 		 & 2.0			& 2.0 \\
Start, UT           & 2008 Jan 08 23:32   	& 2008 Feb 09 07:42		& 2008 Feb 10 08:36 \\
Air mass                        &      1.49		& 1.02			& 1.00 \\
Dichroic split, \AA           &          5300		& 5400			& 5400 \\
Flux standards:		& F66, F110	& F34, F67, G191B2B	& F34, F67, G191B2B \\
Blue:   	wavelength range (\AA)& 	3150-5350	& 3750-5400	&	3650-5350 \\	
	Dispersion (\AA/pixel)         & 4.88 		& 2.63 &			2.63 \\
	Spectral FWHM (\AA)  & 12.1		& 5.4		&	5.4\\
	Scale along slit (arcsec/pixel)  &     0.40 	& 0.72		&	0.72 \\
Red:       wavelength range (\AA)	& 5160-10050	& 5160-7680	& 5650-7740 \\
	Dispersion  (\AA/pixel)          &5.44 		& 2.33		&	2.35 \\
	Spectral FWHM  (\AA)     & 12.8             	& 7.3 		&	7.3 \\
	Scale along slit   (arcsec/pixel) &    0.45	&	0.76		&	0.76 \\
\hline
\end{tabular}
\caption{Spectroscopic observations. Both systems used double spectrographs with red and blue beams separated by a  dichroic beamsplitter, so properties of each spectrum are listed. Wavelength ranges are those of useful data, where both signal-to-noise ratio and quality of the wavelength solution were high.  The two dichroics have very similar properties. [O III] was recorded in both red and blue channels for 2 of the 3 data sets, but H$\beta$ only in the blue. }\label{tab:spectra}
\end{table*}

Wavelength calibration was performed using standard lamps at each telescope. For the Lick red spectrum, the Ne lamp lacks lines shortward of 5852 \AA , so we supplemented this with $\lambda 5577$ night-sky emission from object data to constrain the fit further. The blue WHT data have the worst wavelength solution, because the CuAr+CuNe lamp has substantial line blending at low dispersion; the rms scatter of individual line wavelengths about the fit was 0.8 \AA, or 0.16 pixels. In the other cases, the line scatter about the adopted fits was 0.11-0.17 \AA, or 0.03-0.05 pixels. The line lamps were measured at the beginning or end of the nights, so night-sky lines were used to check for zero-point drifts. In particular, the wavelength scale of the WHT red spectrum requires an offset of -22 \AA. The 2D spectra (object and standard star) were rebinned to linear wavelength scales, confined to the regions where the wavelength solution was well determined. Nyquist `ringing' occurs at the few per cent level for pixels adjacent to [O III]$\lambda$ 5007 emission after wavelength rebinning.

Sky subtraction used a 3rd-order Chebyshev function fit to sections of the slit free from significant galaxy light and any obvious emission at the wavelengths corresponding to H$\alpha$ or [O III]$\lambda\lambda$ 4959 5007 \AA, including a small section between IC 2497 and Hanny's Voorwerp.

Flux calibration used available standard stars. For the WHT, two standard star observations were used although one was only useful in the red. Three stars were used for the first Lick data set and two on the second Lick night. In this latter case, response curves from the two stars agree well in shape but only at 50\% level in intensity. Each of the standard stars has calibrated flux data at 50~\AA\  intervals, except in the deep-red telluric bands, so the sensitivity curves are well constrained; individual flux points scatter about the fit by 0.2 magnitude. A grey shift was thus introduced to match the mean levels for all observations, reducing this scatter to 0.03 magnitude.

\begin{figure}
\includegraphics[width=0.25\textwidth,angle=90]{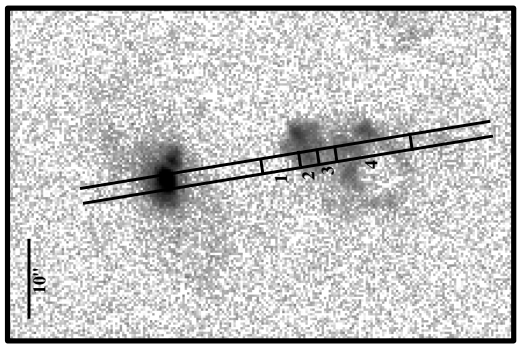}
\caption{Slit position for both WHT and Lick data plotted on a H-alpha image with non-linear scaling in intensity to show detail in both IC2497 and the Voorwerp. The regions labelled 1,2,3 and 4 correspond to the `zones' in table \ref{table:iongrad}.}\label{fig:slit}
\end{figure}

The merged blue and red WHT spectra are shown in Figure \ref{fig:Voorspec}. This represents the flux summed over a region of slit 15-36 arcsec from the nucleus of IC 2497, encompassing the brightest emission from Hanny's Voorwerp. We use this region in assessing overall spectroscopic properties. Although the Lick spectra are not as sensitive as those obtained with the WHT, they have higher spectral resolution and thus give tighter limits on line widths. They are crucial in fully resolving the density-sensitive [S II]$\lambda\lambda$ 6717,6731~\AA~ doublet.

As a further check, we compare the flux obtained from each of the five spectra where the line could be measured. They give a mean integrated flux of $5.7 \times 10^{-14}$ erg cm$^{-2}$ s$^{-1}$ with rms scatter of 23\%.  The flux ratio of the $\lambda$5007 to $\lambda$4959 lines gives an additional check on the errors since the flux ratio should always be 2.93, from statistical weights of the energy levels involved. The measured mean value is 2.92, with rms scatter 10\%.

The spectrum is dominated by a series of emission lines (Table \ref{tab:lines}), with [O\textsc{iii}] at a rest wavelength of 5007\AA~by far the most prominent. Using the higher-resolution Lick data, comparing with the peak of [O III]$\lambda$5007 emission from IC 2497, and intensity weighting along the slit, we derive a mean intensity-weighted redshift for Hanny's Voorwerp which is $269\pm20\mathrm{km~s^{-1}}$ less than that measured for IC 2497. This suggests a genuine physical association between the Voorwerp and IC 2497, rather than a line-of-sight projection effect. The emission spectrum and the accompanying continuum are so dominant that we find only
indirect hints of a population of stars within Hanny's Voorwerp (see section \ref{sec:contm}).

We can use the SDSS $g$ image to estimate the total [O III]
$\lambda 5007$ flux from the object for comparison with the small region sampled by the spectrograph slits. We use the energy zero points for the SDSS system from \citet{Fukugita}, and incorporate the line wavelengths and equivalent widths from the spectra. The total flux we derive in the $\lambda5007$ line is $3.2 \times 10^{-13}$ erg cm$^{-2}$ s$^{-1}$, accounting for a fraction \textbf{$\sim 0.5$} of the total $g$ intensity. Of this, the deeper INT $g$ image shows that a fraction 0.236 of the intensity of
the main body, without outlying patches, falls within the 2-arcsec spectroscopic slit location, so the images give a flux within the spectroscopic slit totalling $4.0 \pm 0.8 \times 10^{-14}$ erg cm$^{-2}$ s$^{-1}$, with the error dominated by the line's equivalent width against the weak continuum at this wavelength. This is consistent {\bf within the errors} with the spectroscopic sum derived earlier.

\section{\emph{Swift} UV/X-ray Data}
\label{sec:swift}
The \emph{Swift} satellite was used to obtain UV and X-ray data toward the Voorwerp\footnote{Archive reference numbers for observations are SW00031116001/2.k }. Observations with the UV/Optical Telescope (UVOT)  and the X-Ray Telescope (XRT) took place for 937s on the 2008 February 8 and 3816 seconds on 2008 February 13. The only strong emission line in the {\emph{Swift} UVW2 filter at $z=0.048$  might be [C\textsc{iii}] at 1909\AA~, but emission from this line is strongly weighted to higher density gas such as that found in AGN broad-line emission regions rather than the low density gas in the Voorwerp (Section \ref{sec:contm}). We also use the X-ray telescope on board \emph{Swift} to search for AGN emission from the larger galaxy or from the Voorwerp itself. 

Observations with the UVOT telescope consisted of a total of 4700 seconds' exposure. As shown in Figure \ref{fig:swiftuv} the Voorwerp is a strong UV source, with $\sim 0.3$ counts per second corresponding, for a flat continuum to a flux of $2.51\times 10^{-16}\mathrm{erg~cm^{-2}}$. The UV continuum image shown in Figure \ref{fig:swiftuv} reveals approximately the same structure seen in [O\textsc{iii}] $\lambda\lambda$ 4959, 5007 emission. In particular, the `bubble' seen to the south-west of centre in the optical image is also evident in UV, although the contrast is much less marked.

\begin{figure}
\includegraphics[angle=90, width=0.48\textwidth]{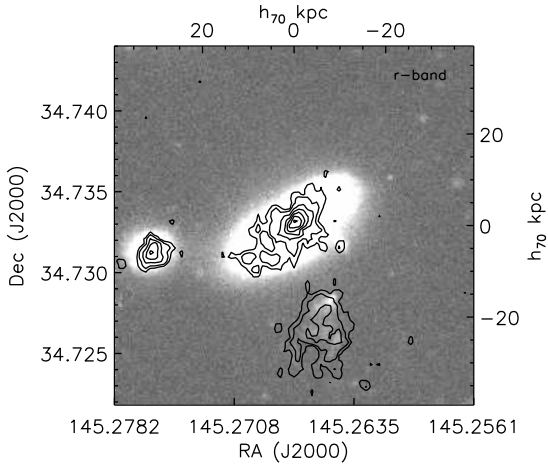}

\caption{Image of the field obtained after 4700 seconds exposure with the UV/optical telescope on \emph{Swift}. The UV data are shown as contours overlaid on a $r$ band image from SDSS. Both IC 2497 (the central source) and the Voorwerp are strong UV sources.}\label{fig:swiftuv}

\end{figure}

IC 2497 - particularly its nucleus -  is also a bright source in the UV, despite the nearby dust lane seen in optical images.  This indicates that the dust lane, while projected near the core, does not actually obscure the central bright part of the galaxy bulge. The apparent spiral arm to the southeast is also distinct in the UV image.

There was no detection of either IC 2497 or the Voorwerp in the X-ray data. Statistics in ``blank-sky" regions confirm that fewer than three counts were obtained from either object in 3700 seconds of integration, giving a count rate of less than 0.001 count $\mathrm{s^{-1}}$. Taking a mean effective area between 2 and 10 keV of 90 $\mathrm{cm^{-2}}$, the sensitivity of the observations was roughly $7.6\times 10^{-14} \mathrm{erg~cm^{-2}~s}$. At the distance of IC2497, this corresponds to a limit of $3.3 \times 10^{41}$ erg s$^{-1}$ between 2-10 keV.

 \section{\bf Physical conditions in the Voorwerp}
 \subsection{Emission-line ratios and diagnostics}


Emission lines provide significant information about the physical conditions in the object and on possible sources of ionization.  For the analysis below we concentrate on the spectrum summed across the brightest region (as in Figure \ref{fig:Voorspec}). 

The density-sensitive [S II] $\lambda 6717/6731$ doublet ratio
is, within the errors, in the low-density limit. Specifically, from the higher-dispersion Lick data for which the lines are fully resolved, the ratio is $1.52 \pm 0.15$; we this derive an upper limit on the density of $n_e < 50$ cm$^{-3}$.

Detection of the [O III] $\lambda 4363$ line provides an estimate of the electron temperature via its ratio with the strong $\lambda 4959, 5007$ lines \citep{Peimbert}. The observed ratio corresponds to a range T$_e$ = 13500 $\pm 1300$ K.

Evidence for internal reddening from the Balmer decrement is equivocal, with errors in the line ratio which are relatively large for such strong lines because we do not have measurements of H$\alpha$ and H$\beta$ on the same detector. The ratio H$\alpha$/H$\beta$=3.2$\pm 0.3$ corresponds to (foreground screen) reddening E$_{B-V} =0.12 \pm 0.10$ for a Milky Way extinction law, assuming an intrinsic  H$\alpha$/H$\beta$ ratio of 2.87 (appropriate for a case B recombination and a temperature of 10000K \citep{Osterbrock}). We do not correct our measured value for internal extinction in our discussion; nonzero extinction would increase the luminosity and slightly decrease the ionization parameter derived, and have the net effect of narrowing the bounds we derive on the ionizing luminosity for the central source.

The most unusual feature of the spectrum of the Voorwerp is the presence of strong emission lines associated with high-ionization species such as He \textsc{ii}$\lambda$4616~\AA~ and [Ne\textsc{v}]$\lambda\lambda$~3346,3426~\AA. We estimate an ionization parameter $U$ following \citet{Penston} and \citet{Komossa}. While the He II/H$\beta$ and [Ne\textsc{v}]/[Ne\textsc{iii}] ratios depend on $U$, they also depend strongly on the shape of
the ionizing spectrum \citep{Komossa}. We thus concentrate on the [O\textsc{ii}] $\lambda 3727$ / [O\textsc{iii}] $\lambda 5007$ ratio, which the models cited find to be more robust. Using an analytical fit to interpolate between models listed by \citet{Komossa}, we find log $U = -2.2$.~Together with the electron density, 
this gives an upper bound on the luminosity of the ionizing source.

 Over a wide range of conditions in ionized nebulae, the ratios [N II]/H$\alpha$ and [S II]/H$\alpha$ scale broadly with abundances. These are both small in Hanny's Voorwerp, the [N II]$\lambda$~6583~\AA~ line in particular suggesting subsolar abundances (crudely $\sim 0.1-0.2 Z_{\odot}$).
 
Several diagnostic line ratios show significant changes with position along the slit,
in the general sense of ionization increasing southward (away from IC 2497). This
is illustrated in 
Table \ref{table:iongrad}  and Figure \ref{fig:iongrad}.~ In particular [Ne\textsc{v}]/[Ne\textsc{iii}], [O\textsc{iii}]/H$\beta$ and He \textsc{ii}:H$\beta$ all increase with distance from the nucleus of IC 2497.

\begin{figure*}
\includegraphics[width=0.8\textwidth,angle=90]{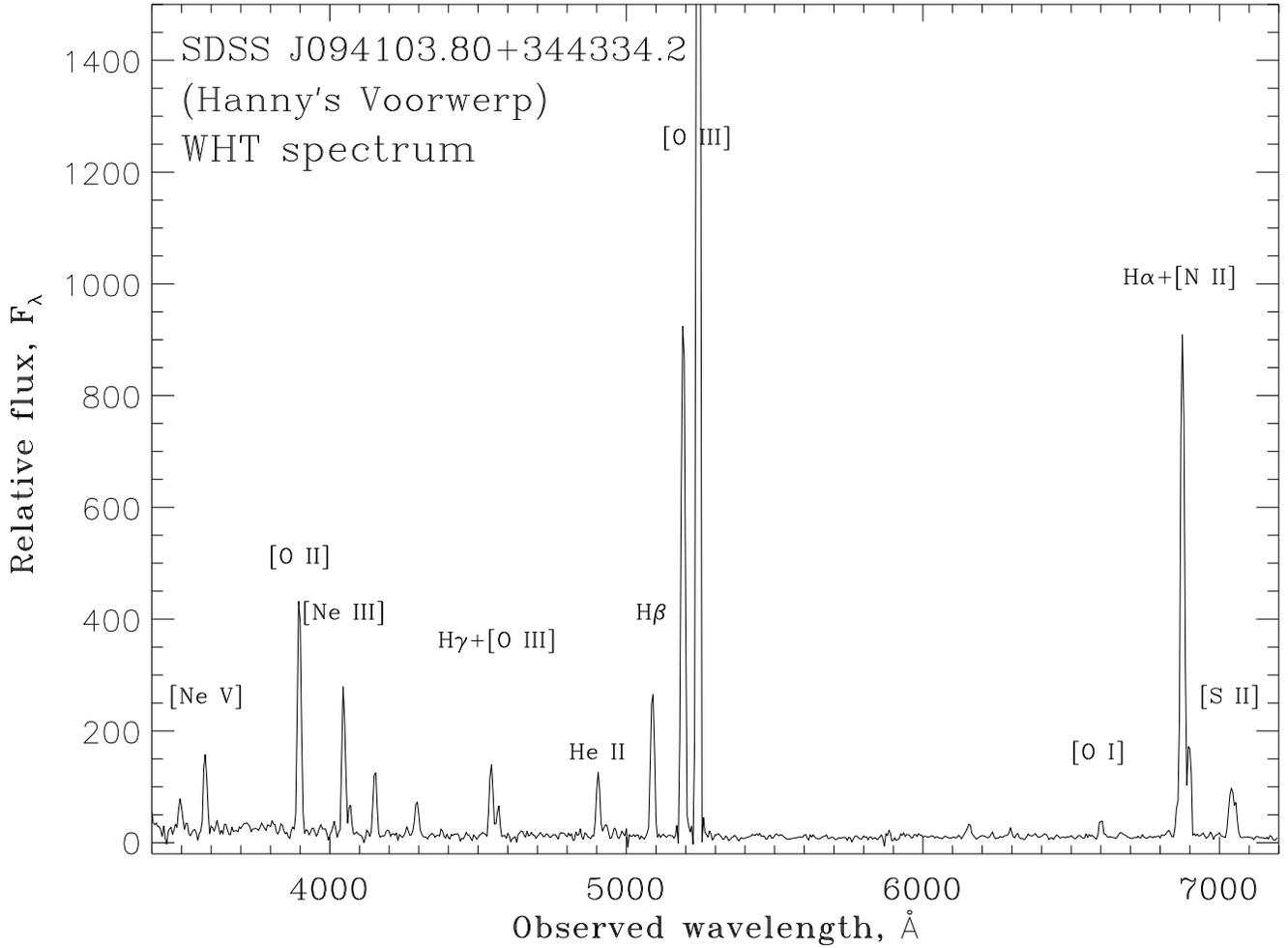}
\caption{Spectrum of Hanny's Voorwerp obtained with the WHT, summed over the slit section 15-36 arcsec. south of the nucleus of IC 2497. The prominent [O\textsc{iii}] 4959, 5007 lines dominate the detected emission, while the presence of [Ne\textsc{v}] and [He\textsc{ii}] lines indicates that the gas is more highly ionized than can be accounted for by starlight. In order to display the fainter lines, the brightest [O\textsc{iii}] line is truncated. Blue and red sections of the spectrum have been merged by resampling to a common wavelength scale, and blended with smoothly varying weights across the range of overlap. }\label{fig:Voorspec}
\end{figure*}

\begin{table}
\begin{tabular}{|cccc}
\hline
Line & Rest & Observed & Ratio with H$\beta$ \\
&  Wavelength (\AA) &  Wavelength (\AA)&\\
\hline

[Ne V] & 3346 & 3496 & $0.2\pm 0.07$ \\
{[Ne V]} & 3426 & 3580 & $0.45 \pm 0.07$ \\
{[O II]} & 3736+3729 & 3897 & $1.54 \pm 0.05$ \\
{[Ne III]} & 3869 & 4046 & $0.83 \pm 0.04$ \\
H$\zeta$ & 3889 & 4067 & $0.17 \pm 0.05$ \\
{[Ne III]}+H$\epsilon$ & 3968+3970 & 4152 & $0.40\pm 0.03$\\
H$\delta$ & 4101 & 4294 & $0.21 \pm 0.03$\\
H$\gamma$ & 4340 & 4544 & $0.48 \pm 0.03$\\
{[O III]} & 4363 & 4568 & $0.12 \pm 0.03$\\
He II & 4686 & 4904 & $0.40 \pm 0.02$\\
H$\beta$ & 4861 & 5088 & 1.00\\
{[O III]} & 5007 & 5243 & $10.5 \pm 1.$\\
He I & 5876 & 6154 & $0.3 \pm 0.02$\\
{[O I]} & 6300 & 6599 & $0.09 \pm 0.02$\\
H$\alpha$ & 663 & 6876 & $3.2 \pm 0.3$\\
{[N II]} & 6583 & 6899 & $0.55 \pm 0.05$\\
{[S II]} & 6717 & 7038 & $0.32 \pm 0.02$\\
{[S II]} & 6731 & 7054 & $0.21 \pm 0.02$ \\
{[S III]} & 9069 & 9505 & $0.3\pm 0.1$ \\
{[S III]} & 9532 & 9999 & $2.0 \pm 0.3$\\
\hline
\end{tabular}
\caption{Measured emission-line ratios from WHT and Lick spectra. These values are averaged over the $\sim 21$ arcsec slice of the Voorwerp summed in Fig.  \ref{fig:Voorspec}, and represent the weighted mean of values from Lick nights 1 and 2 and WHT data (weights 1:2:3). Errors reflect the scatter in the independent measurements when a line was detected in multiple observations, and are otherwise estimated for ([Ne V] and [S III]) from the line intensity and local noise level..  The [O \textsc{iii}]  $\lambda 5007$ line has a mean surface brightness of $1.4\times 10^{-15} \mathrm{erg~s^{-1}\left(cm^2~s~arcsec^2\right)^{-1}}$ within this region.}\label{tab:lines}
\end{table}

\begin{figure}
\includegraphics[width=0.35\textwidth,angle=90]{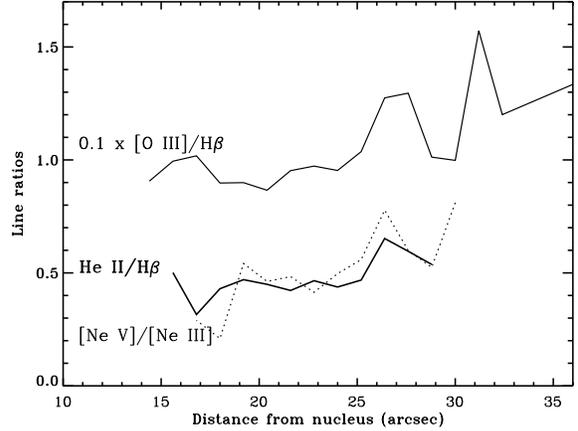}
\caption{The relationship of three emission-line ratios, each associated with the ionization fraction, with distance from the nucleus of IC 2497. [O\textsc{iii}] refers to [O\textsc{iii}]$\lambda$~5007~\AA~,[Ne\textsc{v}] to [Ne\textsc{v}]$\lambda$~3426~\AA, and [Ne\textsc{iii}] to [Ne\textsc{iii}]$\lambda$~3869~\AA. All change with distance from IC 497 in the sense indicating increasing ionization level farther from IC 2497, suggesting that the source of the ionization of the Voorwerp has something to do with the neighbouring galaxy (albeit, perhaps, with the details being complicated).}\label{fig:iongrad}
\end{figure} 
 
\begin{table*}
\begin{tabular}{|c|c|c|c|c|c|c|c|c|}
\hline
Zone&Distance (kpc)&[N\textsc{ii}]/H$\alpha$&[S\textsc{ii}] 6717\AA/H$\alpha$&[S\textsc{ii}] 6731\AA/H$\alpha$&[S\textsc{iii}]/H$\alpha$&[O\textsc{iii}]/H$\beta$&He\textsc{ii}/H$\beta$&[Ne\textsc{v}]/[Ne\textsc{iii}]\\
\hline
1&13-16&0.31&0.15&0.08&0.16&9.7 &0.34&---\\
2&16-19&0.25&0.12&0.08&.39&10.0& 0.34&0.79\\
3&19-22&0.15&0.07&0.06&0.58&9.7& 0.42&0.65\\
4&22-31&0.09&0.07&0.04&0.64&10.7&0.46&0.39\\
\hline
Nucleus&0.8&1.15&0.27&0.27&---&3.6(1.0)&---&---\\
\hline
\end{tabular}
\caption{Emission-line ratios for four averaged positions across Hanny's Voorwerp and for the nucleus of IC 2497. The regions used are indicated on Figure \ref{fig:slit}. Except for [S\textsc{ii}] where we give both lines, where appropriate we refer to the stronger line of a pair so that [O\textsc{iii}] represents 5007\AA, [N\textsc{ii}] 6583\AA~ and [O\textsc{i}] is 6300\AA. [Ne \textsc{iii}] and [Ne\textsc{v}] are the single lines at 3969 and 3426\AA~ respectively. The error bars given in parentheses for the nucleus indicates the difference expected from subtracting a plausible range of stellar populations, which is significant for H$\beta$ because of the relatively strong and uncertain correction for underlying absorption.}\label{table:iongrad}
\end{table*}

\subsection{\bf Continuum: recombination, two-photon emission, and other sources}
\label{sec:contm}

Continuum radiation is evident in the spectra, especially in the blue, 
and the intensity of the {\it Swift} UV image suggests that this part of the spectrum is also dominated by the continuum. We consider here its spectral shape and possible constituents. We combine imaging and spectroscopic results, all scaled to encompass the region summed along the slit for the spectrum shown in figure \ref{fig:Voorspec} (2 arcsec. wide, 15-36 arcsec. S of the nucleus of IC 2497 along PA 8$^{\circ}$). For the spectroscopic points for both WHT and Lick data means in windows free of strong emission lines were used with errors obtained by combining the internal error of the mean with an external 10\% flux-scale error. From images, we use the continuum $\lambda$6573 image from the KPNO 2.1m telescope and the longer exposure in Bessell $R$ (which excludes H$\alpha$ at this redshift) from the SARA 0.9m. We also include the {\it Swift} UVOT measurement. The UVOT passband includes [C III]$\lambda 1909$, so we assign error bars reflecting the range of [C III] : [O III] $\lambda 5007$ ratios seen in ionization cones from Seyfert galaxies with similar ionization levels (Evans et al. 1999).

The equivalent width of H$\beta$ is $360\pm20$ \AA ~ in the emitted frame. This means that the continuum contributions from recombination and two-photon decay from the metastable 2 $^2$S$_{1/2}$ state of H\textsc{i} are not negligible. The equivalent width of H$\beta$ against the recombination (free-free plus bound-free) continuum is 1350\AA~ at the derived electron temperature (De Robertis \& Osterbrock 1986), so that slightly more than a quarter of the observed continuum near H$\beta$ comes from the plasma. We evaluate these contributions using the analytical expressions from Ferland (1980), Osterbrock \& Ferland (2006), and Nussbaumer \& Schmutz (1984). We assume a helium abundance of 0.08 by number, and neglect He$^+$. The two-photon continuum is scaled to conform to the low-density limit with no collisional de-excitation. The sharp Balmer jump is smoothed in practice by the 
pseudocontinuum produced by the confluence of high-order Balmer emission lines, which blends
smoothly into the Balmer continuum. We have approximated this effect in Figure \ref{fig:contm} based on
spectrophotometry of the planetary nebula Jonckheere 900 and the Seyfert galaxy NGC 4151, obtained using the 2.1m telescope on Kitt Peak \citep{Keel87}. These objects bracket the line widths seen in Hanny's Voorwerp; we have logarithmically interpolated the pseudocontinuum in line width, obtaining a shape which is roughly linear in flux between 3646 and 3927 \AA.

As shown in Figure \ref{fig:contm}, the nebular continuum is a significant fraction of the total in this object and most of the emission just shortward of the Balmer jump can be attributed to it. However, the optical emission, roughly flat in $F_{\lambda}$ longward of 5500 \AA, must come from processes other than those that produce the continuum.  Since the normalization of the free-free continuum is set directly from the H$\beta$ equivalent width with error $\pm 6$\%, this excess continuum in the optical data is detected at high significance. Most of the flux measured near 2000 \AA~ is also greatly
in excess of the two-photon continuum and comes from other sources. The longer-wavelength continuum could plausibly represent direct starlight. However, the slope of this residual continuum from the optical to mid-UV is very steep (roughly $\lambda^{-3}$), which may suggest a scattered-light component. Such a contribution should have a strong polarization signature. Given the errors, it is not
clear whether we detect any excess from such additional sources near the Balmer jump at 3646 \AA ;  the entire continuum flux in this region may be accounted for from two-photon emission, Balmer continuum, and the confluence of higher-order Balmer emission lines.  Within the signal-to-noise ratios of the UV and spectroscopic data, we see no differences between the spatial distributions of continuum and line radiation along the spectroscopic slit.

\begin{figure}
\includegraphics[width=0.35\textwidth,angle=90]{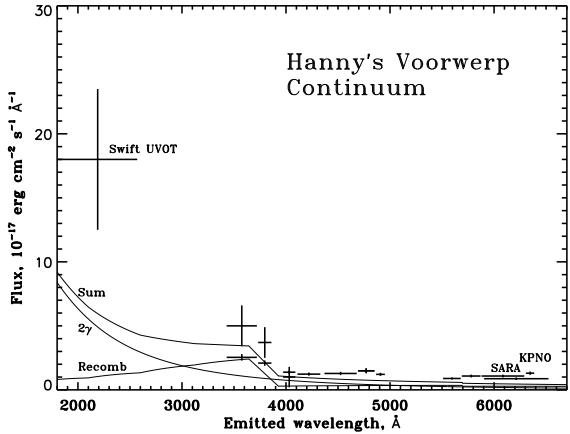}
\caption{The continuum from Hanny's Voorwerp. Binned regions of the spectra between strong emission lines are combined with imaging results. The error bars on the UV point from \emph{Swift} reflect maximum and minimum corrections for [C III]/ $\lambda$ 1909 emission. The curves show the contribution of nebular continuum emission (`Recomb'), two-photon emission (`2 gamma'), an empirical approximation for the pseudocontinuum of blended high-order Balmer emission lines between 3646-3950 \AA\  ,  and the sum of all these components.. Scaled to match the observed equivalent width of H$\beta$, these sources dominate the blue peak in the continuum, but fall short by about a factor 4 in the red and likewise leave much of the UV continuum unaccounted for. These regions may thus include contributions from imbedded starlight or dust scattering of radiation from the ionizing source. }\label{fig:contm}
\end{figure} 

\subsection{\bf Velocity structure}

Significant velocity structure appears in several emission lines, and in both sets of spectral
data. This is best shown in [O\textsc{iii}] $\lambda$ 5007 in the Lick data, which has higher spectral resolution
than the WHT spectrum. Figure \ref{fig:o3dynamics} shows the velocity offset from [O\textsc{iii}] $\lambda$ 5007 in the nucleus of IC 2497, compared to that in the Voorwerp. Errors were estimated following Keel (1996), with a floor of 10 km s$^{-1}$ (corresponding to 0.07 pixel) from pixel centroiding. The peak-to-peak amplitude of this velocity slice is about 90 km s$^{-1}$. Our slit location samples the edge of the `hole' which is a prominent feature in our images; the intensity peaks at 23 and 27 arcsec seen in Figure \ref{fig:o3dynamics} lie along its rim. It may be significant that this region has the most negative radial velocities we observe.

\begin{figure}
\includegraphics[width=0.35\textwidth,angle=90]{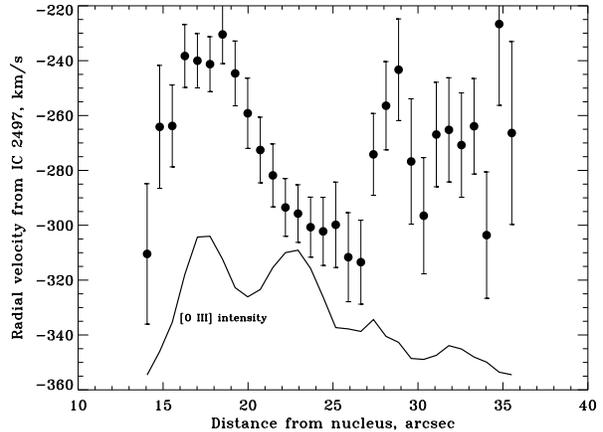}
\caption{Velocity structure in [O III] emission along the spectrograph slit position, with the spatial scale measured from the nucleus of IC 2497. The velocity scale is also centred on [O III] emission from the galaxy nucleus.  Th lower trace shows the intensity of [O III] at each point on the slit. The modest velocity amplitude and lack of consistent correlations between emission peaks and either extreme velocities or gradients argue against an important role for shock ionization. The intensity peaks at 23 and 27 arcseconds lie in the region associated with the crossing of the rim of the `hole' which is prominent in direct images.}\label{fig:o3dynamics}
\end{figure} 

\section{Physical conditions in the nucleus of IC 2497}

Spectral lines were also detected toward the nucleus of IC 2497. Of particular interest are H$\alpha$, [O\textsc{i}]$\lambda$ 6300~\AA, [O\textsc{ii}]$\lambda\lambda$ 3736, 3729~\AA~ and [O\textsc{iii}]$\lambda\lambda$~4959,5007~\AA, from which detections we are able to confirm that the galaxy is in fact a low ionization nuclear emission-line region (LINER) active galactic nucleus with, taking into account the underlying stellar absorption, [N\textsc{ii}]/H$\alpha$=$0.8\pm0.15$ \citep{Heckman}. The [O\textsc{ii}]$\lambda$~3727/[O\textsc{iii}]$\lambda$~5007 ratio is $1.33\pm0.16$. Other important ratios are given in table \ref{table:iongrad}.

Detection of [S\textsc{ii}] $\lambda\lambda$ 6717,6731allows us to calculate the ionizing flux. An ionization parameter for the circumnuclear gas of $10^{-3.2}$ is found. From the measured [S\textsc{ii}]$\lambda$~6717/ [S\textsc{ii}]$\lambda$ 6731 ratio of $1.02\pm0.05$ we derive an electron density of $\sim 560\pm150 \mathrm{cm^{-3}}$ for the centre of IC 2497, similar to values found for the narrow line region in Seyfert galaxies \citep{Peterson, Bennert06, Osterbrock}. As often happens in analysis of LINERs, correction for the underlying starlight is a particular issue for H$\beta$.  Most of the stellar features are well fitted by a template based on elliptical galaxies \citep{Kennicutt}, but an absorption blend around 3850~\AA\  is deeper in the template than in IC 2497. This may hint at a population of younger stars. The effect of such a population would be to simultaneously decrease the [O\textsc{iii}]/H$\beta$ ratio, decrease the reddening needed to account for the H$\alpha$/H$\beta$ ratio, and increase the implied ionization level if the [O\textsc{ii}]/[O\textsc{iii}] ratio were to be corrected by an appropriate amount to match the H$\alpha$/H$\beta$ ratio. In view of these uncertainties, we assign a relatively large error to the [O\textsc{iii}]/H$\beta$ ratio and do not attempt to correct for reddening. Such a correction would, in any case, reduce the derived ionization parameter, so this is a conservative approach in evaluating the level of nuclear activity ionizing the gas.

The line ratios in both IC 2497 and Hanny's Voorwerp are illustrated in the `BPT diagram' \citep{BPT} shown in Figure \ref{fig:BPT}. The trend to increasing ionization with greater distance from IC 2497 is clearly seen in the data for the Voorwerp, although all the points fall in the part of the BPT diagram defined by the Seyfert regime, whereas IC 2497 lies near the boundary between the parts of this diagnostic diagram associated with LINERs and Seyfert nuclei.

\begin{figure*}
\includegraphics[width=0.4\textwidth,angle=90]{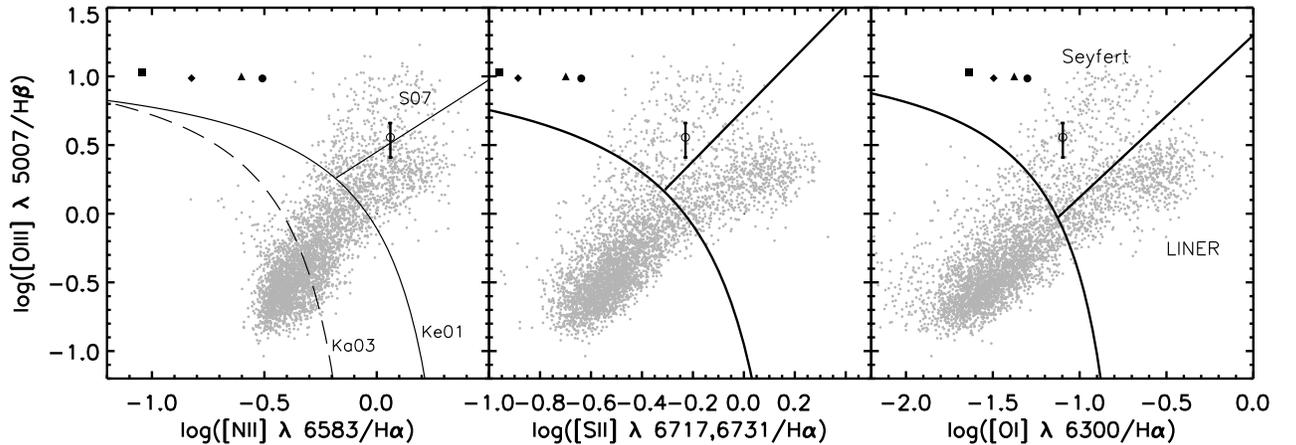}
\caption{BPT diagram for IC 2497 (empty circle) and for four zones in Hanny's Voorwerp. In order, moving away from IC 2497, they are centred on linear projected separations of 13 (filled circle), 17 (square), 20 (triangle) and 27 (diamond)kpc. In the first panel, star-forming galaxies are delineated by the dashed line (\citet{Kauffmann}, or Ka03). Galaxies between this line and the solid (\citet{Kewley}, or Ke01) line have contributions from both AGN and star formation, whereas those beyond this line are pure AGN. The straight line divides Seyfert galaxies from LINERs \citep{Schawinski}. A selection of galaxies extracted from the SDSS is shown in grey for the purposes of comparison.}\label{fig:BPT}
\end{figure*}

\section{Discussion}
\label{sec:discuss}
\subsection{Ionizing the Voorwerp : Photoionization vs Shocks}

Our observations suggest that Hanny's Voorwerp is a low-density gas-rich object, illuminated by a hard ionizing radiation field impinging on the gas. The source of the gas may be IC2497 itself, or the Voorwerp may be a independent dwarf galaxy.  This latter case is suggested by the low derived metallicity, similar to those found for dwarf galaxies by \citet{Tremonti}.

 Gas can be highly ionized either through photoionization by
 a continuum extending to high energies (soft X-rays in this case, since Ne$^{4+}$ has
 an ionization threshold near 100 eV) or fast shocks. The shock interpretation is difficult to sustain in
 this instance, for several reasons. Shock velocities of $\mathrm{400 km s^{-1}}$ are needed to produce
 strong He II and [Ne V] emission \citep{Dopita}, and such velocities are far beyond the radial velocity range of 90 
 km s$^{-1}$ observed here. The lack of a systematic correlation between either
 extreme velocities or velocity gradients and [O III]$\lambda$~5007~\AA~ surface brightness (Figure \ref{fig:o3dynamics}) argues against large-scale shocks as the means of energy input. Finally, shock models give relations among electron
temperature, as measured via the [O III] $\lambda 5007 / \lambda 4363$ ratio, and ionization
indicators such as He II/H$\beta$, which require much higher electron temperatures than we see
in this case \citep{Evans}, typically $\mathrm{T_e\approx 2 \times 10^4K}$.

Although imaging of the Voorwerp at a wide range of wavelengths (including UV imaging and $g$, $r$ and $i$ bands) reveals the presence of a bubble-like structure which is $\sim5 \mathrm{kpc}$ across, and might represent a kind of expanding Str\"omgren sphere, powered by a heavily obscured central source, nothing in the available data suggests such a source.  Instead, we must look for a source of ionization external to the Voorwerp itself. It has a similar redshift to IC 2497, suggesting a genuine physical association. Moreover, the increase in ionization level observed across the Voorwerp, decreasing with distance from IC 2497 supports the hypothesis that the neightbouring galaxy is the direct or indirect source of the ionization. 

One possible counterpart to the Voorwerp which is the result of the action of a jet is Minkowski's object (MO), a blue object near NGC 541 within galaxy cluster Abell 194 \citep{Minkowski, van Breugel, Croft}. There is strong evidence that star formation observed in MO was triggered by a radio jet from NGC 541; and we can thus compare this exotic object with Hanny's Voorwerp to look for evidence of a similar origin. Without a detailed search for such a jet in the IC 2497 system it is difficult to say for certain, but there are important observed differences between MO and the Voorwerp. In particular, optical emission from MO is dominated by [O\textsc{ii}] and H$\alpha$, whereas in the Voorwerp both of these lines are much weaker than the main [O\textsc{iii}] 4959, 5007 line.  MO also exhibits bright continuum emission, whereas the emission lines are clearly dominant in the Voorwerp spectrum. These results suggest that the source of the Voorwerp's ionization is different from that in MO; not hot stars, but something else.

It is also unlikely that the energy input results from direct interaction with outflows from IC 2497, such as radio jets. \citet{Jozsa} report the detection of such a jet associated with the galaxy, but as noted above, shocks from such an interaction would also have to be much faster than the observed velocity range of the gas to account for the high level of ionization. 

\subsection{An AGN in IC2497?}

Having ruled out shocks and interaction with radio jets as the cause of the ionization of the Voorwerp, we next consider a possible AGN in IC2497.  This hypothesis is supported by the observed strength of high-ionization species such as He II and [Ne V], which distinguish this object from typical star-forming regions. The best match to these emission-line ratios (as seen in Figure \ref{fig:BPT}) occurs for gas under conditions  similar to those seen in the narrow-line regions of active galactic nuclei \citep{Leipski, McCarthy}, particularly the
distant gas forming the `extended emission-line regions' tens of kiloparsecs in size seen
around some QSOs and radio galaxies (see summaries by \citet{Fu}, and \citet{StocktonNewAR}), with typical [O III]$\lambda$~5007~\AA~ luminosity exceeding $10^{42}\mathrm{erg s^{-1}}$. They are most prevalent accompanying radio-loud quasars but are not structurally related to either the radio sources or host galaxies.

We now constrain the strength of any AGN in several ways: obtaining an upper limit from the lack of an X-ray detection, both upper and lower bounds from the observed emission-line spectrum, and the level of possibly absorbed AGN radiation from the \emph{IRAS} observations discussed in section \ref{sec:IC2497}.

A lower limit to the required energy input to the gas comes from straightforward energy balance - the number of ionizations and recombinations must match, and the rate of emission of ionizing photons must be at least sufficient to power the observed recombination lines. The integrated H$\beta$ luminosity of the Voorwerp is 
$1.4 \times 10^{41}$ erg $\mathrm{s}^{-1}$. For typical nebular conditions and a flat ionizing continuum, one in 12.2 recombinations cascades through the H$\beta$ transition and one in 9.1 for H$\alpha$ (table 4.4 in \citet{Osterbrock}). The fraction of the ionizing luminosity (between H and He ionization edges) reprocessed into line emission depends on both the optical depth (making the derived luminosity a lower limit) and covering fraction. In our deepest $g$ image, the emission subtends approximately 38$^\circ$ about the nucleus of IC 2497, which would correspond to a covering fraction of $\sim0.03$ if it is comparably deep along the line of sight. For a flat ionizing continuum ($F_\lambda \propto \nu ^{-1}$), this gives a required ionizing luminosity $>1.0 \times 10^{45}$ erg
 $\mathrm{s}^{-1}$; the X-ray luminosity is comparable for this continuum slope.

Since we have an upper limit to the electron density, we can use the ionization parameter in the gas to provide an upper limit to the incident continuum flux and hence luminosity. For ionization parameter $U = 0.006$ and $n_e < 50$ cm$^{-3}$, the local density of ionizing photons will be $< 0.32$ cm$^{-3}$. Using the mean projected separation of the Voorwerp from the core of IC 2497, 20kpc, as the distance the ionizing source must have an isotropic output of  $Q_{ion} < 9.5 \times 10^{56}$ s$^{-1}$. For a flat continuum shape, this corresponds to $L_{\mathrm{ion}} < 3.2 \times 10^{45}$ erg $\mathrm{s}^{-1}$, and again a comparable X-ray output would be expected. These two emission-line arguments thus bound the required ionizing luminosity in the range $1-3 \times 10^{45}$ erg $\mathrm{s}^{-1}$.

We can place a limit on any nuclear activity in IC 2497 with the \emph{Swift} X-ray data, described in section \ref{sec:swift}. Assuming an unabsorbed, AGN-like, ($\nu^{-1}$) spectrum between 2-10 keV, these data rule out relevant AGN luminosities; for a flat, unobscured continuum the limit derived from our \textit{Swift} observations  is more than 3dex below the required luminosities.  This conclusion holds unless, along our line of sight, most of the flux up to 5 keV is absorbed (more specifically, using the XSPEC web tool at the HEASARC site, this means equivalent $N_\mathrm{H} < 10^{24}$ cm$^{-2}$).

 The key issue is therefore whether IC 2497 could host a powerful AGN which remains active and luminous, but is so deeply obscured as to elude our observations so far.

The far-IR data from the IRAS survey (see section \ref{sec:IC2497}) suggest a FIR luminosity of $1.5\times10^{44}\mathrm{erg~s^{-1}}$, an order of magnitude less than the energy requirements we find for ionizing luminosity.  This observed value, which applies to the integrated flux of the galaxy, includes any contribution from the disk of what is a very luminous spiral on top of the AGN component. In comparing the total FIR output to the isotropic ionizing flux, we implicitly assume a geometry in which most of the radiation is intercepted by some thick, roughly toroidal structure, suggested by known AGN in which strong obscuration shapes the spatial extent of the escaping radiation \citep{Tadhunter, Mason}. Other structures are possible; if the obscuring material is a thinner annulus seen edge-on, or patchy and of small covering fraction but blocking our line of sight, the amount of energy absorbed could be proportionally smaller. Such material would have to satisfy the column density constraints from X-ray observations.

However, this simple picture is unlikely to apply here as the nuclear gas in IC 2497 sees very little ionizing radiation. The emission spectrum from the nucleus of IC 2497 is quite representative of the LINERs often found in early-type spirals, and well explained by a power-law continuum of low ionization parameter log $U \approx -3.5$. Obscuration strong enough to block our line of sight and thus hide the core seems unlikely without raising the ionization level of circumnuclear material beyond the observed value. In addition, the ionization cones seen in galaxies such as Cygnus A and NGC 1068 are smaller-scale features which surround the nucleus rather than illuminating distant patches of ionisation such as we see here. 

The region of gas which is seen in emission in a LINER is typically within 0.5kpc from the nucleus \citep{Prieto}, consistent with the region included in the seeing disk of the nucleus for our spectra. If this distance is an appropriate estimate for conditions in IC 2497, then  gas which is $\sim40$ times closer to the central ionizing source than the Voorwerp must be seeing an ionizing flux less than half of that seen by the more distant gas. In order to reconcile these observations, we would be forced to postulate some geometry which allows ionizing radiation to escape from the galaxy \emph{only} through a channel some 20 degrees in half-angle, without encountering a significant density of the galactic interstellar medium.  

Furthermore, the Swift/XRT observation also rules out a luminous, Compton-thick AGN 
currently residing in IC 2497. Consider a heavily obscured AGN with an 
intrinsic column density of $1\times 10^{24} \mathrm{cm^{-2}}$, i.e. Compton-thick, and an 
unabsorbed luminosity of $10^{45} \mathrm{erg~s^{-1}}$ which  matches the ionizing 
luminosity required to explain the ionization in Hanny's Voorwerp. While most of 
intrinsic flux is absorbed, Levenson et al (2006) find that  
approximately 1\% of the continuum's intrinsic flux is detected in 
reflection for 7 Compton-thick AGN studied with the Chandra X-ray 
Observatory. If IC 2497 hosts a $10\times 10^{45} \mathrm{erg~s^{-1}}$ Compton-thick AGN, the 
expected observed 0.2--10 keV flux is $1.72\times 10^{-12} \mathrm{erg~s^{-1}~cm^{-2}}$ and PIMMS 
would predict 130 Swift/XRT photons in 3700 seconds.  
Therefore, either the reflection fraction in IC 2497 is much smaller 
than that indicated by Levenson  et al. (2006) or IC 2497 does not currently 
host a sufficiently powerful heavily obscured AGN. What these data alone do not 
rule out is either a moderately obscured AGN in IC 2497 or 
even something similar to NGC 1068 which is often regarded as the 
prototypical example of a Seyfert 2 nucleus. Harder x-ray observations will be necessary to rule out this latter case.

Thus, from our knowledge of the properties of IC 2497, using observations from the infra-red through to the x-ray, it is difficult to identify  a present-day AGN as the source for the high levels of ionization seen in the Voorwerp, and this leads us to consider an alternative hypothesis. 

\subsection{A quasar light echo?}

In the absence of an ionizing source, we conclude that the Voorwerp was ionized  by a source which is no longer active. We hypothesise that IC 2497 underwent an outburst, reaching quasar luminosities, and that we see material which lies close to the light-echo (or constant time-delay) ellipsoid \citep{Couderc} and is illuminated and ionized by this prior outburst. 

The first astronomical detection of a light echo, around Nova Persei 1901, was described by \citet{Kapteyn}. This discovery has been followed by the discovery of simple scattering echoes from - most famously - SN 1987A, the eruptive variable V838 Monocerotis \citep{Bond}, and from more distant extragalactic supernov\ae~ (e.g. \citealp{Rest}). Light echos have recently been exploited to measure the spectra of historical supernov\ae, and deduce their spectroscopic classifications \citep{Rest2}. If our hypothesis is correct, the Voorwerp represents the first detection of the phenomenon with a source that lies on galactic rather than stellar scales. 

The separation of the Voorwerp from IC 2497 is between 45,000 - 70,000 light years, depending on the angle of projection. For a true light echo, as grains are forward scattering, the most favourable scattering geometries for UV dust reflection will place the Voorwerp in front of IC 2497. This suggests that an outburst, or perhaps the end of a longer luminous phase, must have taken place $\sim 10^5$ years ago (referred to the epoch at which we observe IC 2497). The use of ``light echo" would be fully consistent with previous usage only for the dust-scattered component which we infer for the UV continuum. The recombination timescale at the low densities we measure is $> 8000$ years, small but not trivial compared to the light-travel times involved, so the observed emission-line response (``photoionization echo") would be more spread in depth than would be the case for pure reflection.

It has long been clear that the AGN population evolves over time (see e.g. \citealt{Boyle,Wolf,Richards}), but it is harder to constrain the timescales on which individual objects undergo change. The connection between AGN and mergers suggests that the subsequently triggered AGN episodes last typically for $10^{8}$ years \citep{Stockton, Bahcall} and may last for up to $10^{9}$ years \citep{Bennert}. The presence of young stellar populations in many quasar host galaxies suggest that their activity is connected to starbursts with a similar timescale of $\sim10^{8}$ years \citep{Canalizo, Miller}. At the other end of the scale, there have been numerous detections of AGN which flare on timescales of years \citep{Storchi,Cappellari}. The timescale we infer for the shutdown of activity in IC 2497, of $\sim10^5$ years, is intermediate between these extremes. Short timescales $\approx10^6$ years have been suggested for episodes of luminous AGN activity both from the distribution of derived Eddington ratios \citep{Hopkins} and statistics of QSO absorption systems at high redshift \citep{Kirkman}.

The lowest redshift quasar in the SDSS DR5 catalogue \citep{Schneider} lies at z=0.08, but this sample systematically excludes systems at lower redshift. Our best comparison is with \citet{Barger}. Taking the 2-8keV luminosity of $10^44$ (a conservative estimate for the flux required to produce the ionization fraction we observe) the local space density of such luminous AGN is no greater than $3\times 10^{-7}\mathrm{Mpc^3}$. This suggests that there should be one such system at a redshift of $z<0.04$, so the presence of such activity in IC 2497, while unusual, is not entirely unexpected.


If the obscuration along the line of sight to the Voorwerp has remained constant, then the AGN in IC 2497 must have undergone either an extremely bright flare or else reached the end of an extended period of high luminosity. In either case, detailed observation of the Voorwerp would enable us to reconstruct the history of the source, probing AGN variability on timescales of $10^5$ years for the first time. This hypothesis suggests further observations which could test it, and, if it is correct, uncover the details of the object's history. We would expect the scattered continuum to be polarized and show broad QSO emission lines in reflection; this spectral signature would be brightest in the UV, possibly within the range of GALEX for such a large and diffuse target. The variation in ionization parameter might trace changes in the ionizing luminosity; measurements of the density across the object could separate density and time effects. The origin of the gas (and scattering dust) in the Voorwerp may have been a dwarf galaxy, probably close enough to IC 2497 to have been tidally disrupted. Near-IR imagery at high resolution may be the best way to search for star clusters from a pre-existing stellar population with minimal interference from the very blue scattered light and the nebular continuum emission.

\section{Conclusion}

We have presented observations of Hanny's Voorwerp, an object first identified through visual inspection of the Sloan Digital Sky Survey as part of the Galaxy Zoo project. The object, near to and at the same redshift as IC 2497, a spiral galaxy, is highly ionized and has a spectrum dominated by emission lines, particularly [O\textsc{iii}] $\lambda\lambda$ 4959,5007, with no sign of any contribution from a stellar component to the Voorwerp itself. Both the Voorwerp and its neighbouring galaxy are strong UV sources, but neither was detected in X-ray observations carried out with the \emph{Swift} satellite. This lack of X-ray detections, and the limits derived from IRAS observations of IC 2497, provides a strong constraint on the luminosity of any ionizing source. We are left with two possible conclusions. Either an AGN in IC 2497 is heavily obscured but still able to ionize the Voorwerp, which extends over almost 20 degrees, or else the ionizing source is no longer present. In the latter case, the Voorwerp represents the first instance of a light echo being seen from a quasar-luminous AGN. In either case IC 2497 furnishes a nearby example of a galaxy which either is, or was shortly before the epoch at which we observe it, a quasar host galaxy.

Detailed further observations, particularly observations in the radio and deep optical imaging, will be required to confirm our hypothesis. However, it is clear that such a light echo would provide an unusual - possibly unique -  opportunity to probe the variation of an AGN on timescales of $\sim10^5$ years, reconstructing its history by observing echoes from different parts of the Voorwerp.

\section*{Acknowledgements}
CJL acknowledges support from the STFC Science in Society Program. WK acknowledges support from a College Leadership Board Faculty fellowship. KS is supported by the Henry Skynner Junior Research Fellowship at Balliol College, Oxford. NB is supported through a grant from the National Science Foundation, (AST 0507450) and the Space Telescope Science Institute, which is operated by The Association of Universities for Research in Astronomy, Inc., under NASA contract No. NAS526555. 

We thank the Lick Observatory staff for their assistance in obtaining our data, and Misty Bentz, Jonelle Walsh and Jong-Hak Woo for obtaining an additional spectrum at Lick Observatory. R. Antonucci provided helpful comments on an earlier draft of this manuscript, and Gary Ferland provided helpful reminders of fine points of nebular astrophysics. Pamela Gay also helped refine the final draft. Our anonymous referee was also responsible for substantial improvements to the reduction and analysis. We acknowledge the use of public data from the \emph{Swift} data archive, and thank the \emph{Swift} operations team for their rapid response to our data request. The William Herschel and Isaac Newton Telescopes are operated on the island of La Palma by the Isaac Newton Group in the Spanish Observatorio del Roque de los Muchachos of the Instituto de Astrof\'isica de Canarias.

This research has made use of the NASA/IPAC Extragalactic Database (NED)
which is operated by the Jet Propulsion Laboratory, California Institute of
Technology, under contract with the National Aeronautics and Space
Administration.

Funding for the SDSS and SDSS-II has been provided by the Alfred P. Sloan Foundation, the Participating Institutions, the National Science Foundation, the U.S. Department of Energy, the National Aeronautics and Space Administration, the Japanese Monbukagakusho, the Max Planck Society, and the Higher Education Funding Council for England. The SDSS Web Site is http://www.sdss.org/.

The SDSS is managed by the Astrophysical Research Consortium for the Participating Institutions. The Participating Institutions are the American Museum of Natural History, Astrophysical Institute Potsdam, University of Basel, University of Cambridge, Case Western Reserve University, University of Chicago, Drexel University, Fermilab, the Institute for Advanced Study, the Japan Participation Group, Johns Hopkins University, the Joint Institute for Nuclear Astrophysics, the Kavli Institute for Particle Astrophysics and Cosmology, the Korean Scientist Group, the Chinese Academy of Sciences (LAMOST), Los Alamos National Laboratory, the Max-Planck-Institute for Astronomy (MPIA), the Max-Planck-Institute for Astrophysics (MPA), New Mexico State University, Ohio State University, University of Pittsburgh, University of Portsmouth, Princeton University, the United States Naval Observatory, and the University of Washington.

\label{lastpage}


\begin{thebibliography}{99}

\bibitem[\protect\citeauthoryear{Adelman-McCarthy et al.}{2007}]{DR6} Adelman-McCarthy et al., 2007, in press, arXiv 0707.3413

\bibitem[\protect\citeauthoryear{Bahcall et al.}{1997}]{Bahcall} Bahcall, J.N., Kirhakos, S., Saxe, D.H. \& Schneider, D.P., 1997, ApJ, 479, 642

\bibitem[\protect\citeauthoryear{Baldwin, Phillips, Terlevich}{1981}]{BPT} Baldwin, J.A., Phillips, M.M. \& Terlevich, R., 1981, PASP, 93, 5

\bibitem[\protect\citeauthoryear{Barger et al.}{2005}]{Barger} Barger, A.J., et al., 2005, AJ, 129, 578

\bibitem[\protect\citeauthoryear{Becker et al.}{1995}]{Becker} Becker, R.H., White, R.L., Helfand, D.J., 1995, ApJ, 450, 559

\bibitem[\protect\citeauthoryear{Best et al.}{2005}]{Best} Best, P.N., et al., 2005, MNRAS, 362, 25

\bibitem[\protect\citeauthoryear{Bennert et al.}{2006}]{Bennert06} Bennert et al., 2006, A\&A, 459, 55

\bibitem[\protect\citeauthoryear{Bennert et al.}{2008}]{Bennert} Bennert et al., 2008, ApJ, 677, 846

\bibitem[\protect\citeauthoryear{Blanton et al.}{2003}]{Blanton} Blanton, M.R. et al., ApJ, 592, 819

\bibitem[\protect\citeauthoryear{Boller et al.}{2002}]{Boller} Boller, Th et al., 2002, MNRAS, 329, 1

\bibitem[\protect\citeauthoryear{Bond et al.}{2003}]{Bond} Bond, H.E., et al., 2003, Nature, 422, 405

\bibitem[\protect\citeauthoryear{Boyle et al.}{2000}]{Boyle} Boyle, B.J., et al., 2000, MNRAS, 317, 1014

\bibitem[\protect\citeauthoryear{van Breugel et al.}{1985}]{van Breugel} van Breugel, W., Filippenko, A.V., Heckman, T, Miley, G, 1985, ApJ, 293, 83

\bibitem[\protect\citeauthoryear{Canalizo \& Stockton}{2001}]{Canalizo} Canalizo, G., \& Stockton, A., 2001, ApJ, 555, 719 

\bibitem[\protect\citeauthoryear{Cappellari et al}{1999}]{Cappellari} Cappellari et al., 1999, ApJ, 519, 117

\bibitem[\protect\citeauthoryear{Couderc}{1939}]{Couderc} Couderc, P., 1939, Annales, d'Astrophysique, 2, 271

\bibitem[\protect\citeauthoryear{Croft et al.}{2006}]{Croft} Croft, S. et al., 2006, ApJ, 647, 1040 

\bibitem[\protect\citeauthoryear{De Robertis \& Osterbrock}{1986}]{DeRobertis}De Robertis, M.M. \&
Osterbrock, D.E., 1986, PASP, 98, 629 

\bibitem[\protect\citeauthoryear{Dopita \& Sutherland}{1996}]{Dopita}Dopita, M.A. \& Sutherland,
R.S., 1996, ApJS, 102, 161

\bibitem[\protect\citeauthoryear{Dunkley et al.}{2008}]{Dunkley} Dunkley, J. et al, submitted to ApJS, astro-ph/0803.0586

\bibitem[\protect\citeauthoryear{Evans et al.}{1999}]{Evans}Evans, I., Koratkar, A., Allen, M., Dopita, M., \& Tsvetanov, Z.\ 1999, ApJ, 521, 531

 \bibitem[\protect\citeauthoryear {Ferland}{1980}]{Ferland} Ferland, G.J., 1980, PASP, 92, 596
 
 \bibitem[\protect\citeauthoryear{Fisher et al.}{1995}]{Fisher} Fisher, K.B. et al., 1995, ApJS, 100, 69

\bibitem[\protect\citeauthoryear{Fu \& Stockton}{2008}]{Fu} Fu, H. \& Stockton, A., 2008,
ApJ,  in the press (astro-ph/0809.1117)

\bibitem[\protect\citeauthoryear{Fukugita et al.}{1995}]{Fukugita} Fukugita, M., Shimasaku, K., \& Ichakawa, T. 1995, PASP, 107, 925

\bibitem[\protect\citeauthoryear{Heckman}{1980}]{Heckman} Heckman, T.M., A\&A, 87, 152

\bibitem[\protect\citeauthoryear{Hopkins \& Hernquist}{2008}]{Hopkins}Hopkins, P.F. \& Hernquist, L., 2008, ApJ submitted), aXiv.org:0809.3789 

\bibitem[\protect\citeauthoryear{Jozsa}{2009}]{Jozsa}Jozsa, G.I.G. et al., Accepted by A\&A, arXiv.org:0905.1851 


\bibitem[\protect\citeauthoryear{Kapteyn}{1902}]{Kapteyn} Kapteyn, J.C., 1902, Astronomische Nachricheten, 157, 201J. 2006, AJ, 132, 2233

\bibitem[\protect\citeauthoryear{Kauffmann}{2003}]{Kauffmann} Kauffmann, G. et al. 2003, MNRAS, 346, 1055

\bibitem[\protect\citeauthoryear{Keel}{1987}]{Keel87} Keel, W.C., 1987, A\&A, 172, 43

\bibitem[\protect\citeauthoryear{Keel}{1996}]{Keel} Keel, W.C., 1996, ApJS, 106, 27

\bibitem[\protect\citeauthoryear{Keel et al. }{2006}]{Keel0313} Keel, W.C., White, R.E., III, Owen, F.N., \& Ledlow, M.J.,2006, ApJS, 106, 27

\bibitem[\protect\citeauthoryear{Kennicutt}{1998}]{Kennicutt} Kennicutt, R.C, 1998, ApJ, 498, 541

\bibitem[\protect\citeauthoryear{Kewley}{2001}]{Kewley} Kewley, L.J., et al., 2001, ApJ, 556, 121

\bibitem[\protect\citeauthoryear{Kirkman \& Tytler}{2008}]{Kirkman} Kirkman, D. \& Tytler, D., 2008, MNRAS (submitted), arXiv.org:009.2277 

\bibitem[\protect\citeauthoryear{Komossa \& Schulz}{1997}]{Komossa} Komossa, S. \& Schulz, H., 1997, A\&A, 323, 31

\bibitem[\protect\citeauthoryear{Leipski et al.}{2007}]{Leipski} Leipski, C. et al, 2007, A\&A, 467, 895

\bibitem[\protect\citeauthoryear{Levenson et al.}{2006}]{Levenson} Levenson, N.A., Heckman, T.M., Krolik, J.H., Weaver, K.A. \& Zycki, P.T., 2006, ApJ, 648, 111

\bibitem[\protect\citeauthoryear{Lintott et al.}{2008}]{Lintott} Lintott, C. \& the Galaxy Zoo team, 2008, MNRAS, 389, 1179

\bibitem[\protect\citeauthoryear{Lonsdale\& Helou}{1985}]{Lonsdale} Lonsdale, C.~J., \& Helou, G.\
1985,
Cataloged galaxies and quasars observed in the IRAS survey, Pasadena:
Jet Propulsion Laboratory (JPL)

\bibitem[\protect\citeauthoryear{Lupton et al.}{2004}]{Lupton} Lupton, R. et al., PASP, 2004, 116 

\bibitem[\protect\citeauthoryear{McCarthy}{1993}]{McCarthy} McCarthy, 1993, ARA\&A, 31, 639

\bibitem[\protect\citeauthoryear{Mason et al.}{2006}]{Mason} Mason et al., 2006, ApJ, 640, 624

\bibitem[\protect\citeauthoryear{Melbourne et al.}{2004}]{Melbourne} Melbourne, J. et al., 2004, AJ, 127, 686

\bibitem[\protect\citeauthoryear{Miller \& Sheinis}{2003}]{Miller} Miller, J.S. \& Sheinis, A.I., 2003, ApJL, 588, L9

\bibitem[\protect\citeauthoryear{Minkowski}{1958}]{Minkowski} Minkowski, R., 1958, PASP, 70, 143

\bibitem[\protect\citeauthoryear{Nussbaumer \& Schmutz}{1984}]{Nussbaumer} Nussbaumer, 
H. \& Schmutz, W., 1984, A\&A, 138, 495

\bibitem[\protect\citeauthoryear{Osterbrock \& Ferland}{2006}]{Osterbrock} Osterbrock, D.E. \& Ferland, G.J., 2006, Astrophysics of gaseous nebul\ae ~and Active Galactic Nuclei, Second edition, University Science Books, Sausalito, California. 

\bibitem[\protect\citeauthoryear{Peimbert \& Costero}{1969}]{Peimbert} Peimbert, M. \& Costero, R., 1969, Bolet\'in de los Observatorios de Tonantzintla y Tacubaya, 5, 3

\bibitem[\protect\citeauthoryear{Penston et al.}{1990}]{Penston} Penston, M.V. et al., 1990, A\&A, 236, 53

\bibitem[\protect\citeauthoryear{Peterson}{2003}]{Peterson} Peterson, B.M., 2003, An introduction to active galactic nuclei, Cambridge University Press, Cambridge. 

\bibitem[\protect\citeauthoryear{Pottasch et al.}{2008}]{Pottasch} Pottasch, S.~R., Bernard-Salas, J., \& Roellig, T.~L.\ 2008, A\&A, 481, 393

\bibitem[\protect\citeauthoryear{Prieto, Maciejewski \& Reunanen}{2005}]{Prieto} Preito, M.A., Maciejewski, W. \& Reunanen, J., 2005, AJ, 130, 1472

\bibitem[\protect\citeauthoryear{Rest et al.}{2008a}]{Rest} Rest, A. et al., 2008a, ApJ, 680, 1137

\bibitem[\protect\citeauthoryear{Rest et al.}{2008b}]{Rest2} Rest, A. et al., 2008b, ApJ, 681, L81

\bibitem[\protect\citeauthoryear{Richards et al.}{2006}]{Richards} Richards, G. et al., 2006, AJ, 131, 2766

\bibitem[\protect\citeauthoryear{Sanders \& Mirabel}{1996}]{SandersMirabel} Sanders, D.B. \& Mirabel, I.F., 1996, ARA\& A, 34, 749 

\bibitem[\protect\citeauthoryear{Schawinski et al.}{2007}]{Schawinski} Schawinski, K., et al., 2007, MNRAS, 382, 1415

\bibitem[\protect\citeauthoryear{Schneider et al.}{2007}]{Schneider} Schneider, D.P. et al., 2007, AJ, 134, 102

\bibitem[\protect\citeauthoryear{Stockton}{1982}]{Stockton} Stockton, A., 1982, ApJ, 257, 33

\bibitem[\protect\citeauthoryear{Stockton, Fu, \& Canalizo}{2008}]{StocktonNewAR} Stockton, A.,Fu, H., \& Canalizo, G. 2008, NewAR 50, 694

\bibitem[\protect\citeauthoryear{Storchi-Bergmann, Baldwin \& Wilson}{1993}]{Storchi} Storchi-Bergmann, T., Baldwin, J.A. \& Wilson, A.S., 1993, ApJL, 410, L11

\bibitem[\protect\citeauthoryear{Tadhunter et al.}{1999}]{Tadhunter} Tadhunter, C.N., et al., 1999, ApJ, 512, L91

\bibitem[\protect\citeauthoryear{Tremonti et al.}{2004}]{Tremonti} Tremonti et al., ApJ, 613, 898

\bibitem[\protect\citeauthoryear{Wolf et al.}{2003}]{Wolf} Wolf, C., Wisotzki, L., Borch, A., Dye, S., Kleinheinrich, M., Meisenheimer, K., 2003, A\&A, 408, 499

\bibitem[\protect\citeauthoryear{York et al.}{2000}]{York} York, D.G., 2000, AJ, 120, 1579

\end{thebibliography}
\end{document}